\documentstyle[aps,prc]{revtex} 

\begin{document} 
\topmargin .0in
\tightenlines
\title{Baryon Resonance Extraction from $\pi N$ Data using a 
Unitary Multichannel Model}

\author{T.P. Vrana\footnote{Present Address: Fisher Scientific Corp.,
Pittsburgh, PA} and S.A. Dytman\footnote{email:dytman+@pitt.edu;
phone:412-624-9244; fax:412-624-9163}} \ \address{Department of Physics 
and Astronomy, University of Pittsburgh, Pittsburgh, PA 15260}\ 
\author{ T.-S. H. Lee } \
\address{Physics Division, Argonne National Laboratory, 
	Argonne, IL 60439} \

\date{July 12, 1999} 

\maketitle

\begin{abstract}  
A unitary multi-channel approach, first 
developed by the Carnegie-Mellon-Berkeley
group, is applied to extract the pole positions, masses, and 
partial decay widths of 
nucleon resonances from the partial wave amplitudes for the transitions
from $\pi N$ to eight possible final baryon-meson states.  
Results of single energy analyses of the VPI group using the most 
current database are used in this analysis.  A proper treatment of
threshold effects and channel coupling within the unitarity constraint
is shown to be crucial in extracting resonant parameters, especially for
the resonance states, such as S$_{11}$(1535), which have decay 
thresholds very close to the resonance pole position. 
The extracted masses and partial decay widths of 
baryon resonances up to about 2 GeV mass are listed and compared with 
the results from previous analyses.  In many cases, 
the new results agree with
previous analyses.  However, some significant differences, 
in particular for the resonances that are weakly excited in $\pi N$
reactions, are found.

\end{abstract}


\tableofcontents


\section{Introduction} 
\label{se:Intro}

	The interest in the study of baryon resonances began many years
ago and led to the important discovery of SU(3) symmetry.  Although
many states were discovered, the quality and scope of the data
limited the analyses.  Interest has grown significantly 
in the last few years because of the prospects for new data of high 
quality at facilities such as the Thomas Jefferson National 
Accelerator Facility, the Bonn synchrotron, and 
Brookhaven National Laboratory.  These are excited states of the nucleon 
with excitation energy of about 0.3 - 1 GeV.  
Since most of these states were discovered in amplitude analyses of 
$\pi N \rightarrow \pi N$ scattering data, they are labeled by 
the approximate mass and the $\pi N$ quantum numbers:
the relative orbital angular momentum L, the total isospin T,
and the total angular momentum J.  For example, the $D_{13}$(1520) 
resonance has a mass
of about 1520 MeV, isospin T=1/2, total angular momentum J=3/2, and 
decays into a L=2 $\pi N$ state.
Information about the various baryon resonances is evaluated and tabulated 
in biennial publications of the Particle
Data Group\cite{PDG}.  In their 1998 publication, they list about 20 N* 
(T=1/2) and 20 $\Delta$ (T=3/2) states.  Some states are well
established in the data and most analyses agree on their 
properties.  On the other hand, some very 
large discrepancies exist between different analyses. 
For example, the extracted full widths for the 
S$_{11}(1535)$ state (A 4* state of PDG\cite{PDG}) are 66 MeV\cite{VPIpin}, 
120$\pm$20 MeV\cite{KH80}, 151$\pm$27 MeV\cite{ManSal}, 
151-198 MeV\cite{Feuster}, and 270$\pm$50 MeV\cite{Cut79}.  
This state has a number of striking 
properties - unusual strong decay width to $\eta N$~\cite{PDG} and 
an unusually flat transition form factor~\cite{FosterHughes} - that have 
made it a very interesting state to
understand.  
Although there are a number of strongly excited states (rated 4* or 3*
by PDG), there are many weakly excited states whose properties are poorly 
determined with the existing data.
The differences between various amplitude analyses originate 
for various reasons, including handling of data, method of parameterizing
nonresonant background, and handling of the many N* decay channels.
We emphasize the fact that the baryon resonances always decay into
a baryon and one or more of various mesons; the existing 
analyses differ from each other significantly in the methods used to
describe this intrinsically multi-channel problem. 
Thus, model dependence in extraction of resonance properties 
makes it difficult to test predictions of theoretical models
with existing data.  With the new experimental facilities, the 
situation will soon be greatly improved when 
more exclusive data for different final 
states, such as $\eta N$, $\pi \Delta$, and $\rho N$, become available.
One of the major goals of the new experiments is to obtain data for
a large number of reactions.  This greatly increases the probability
of seeing new resonances, e.g. those that couple weakly to the 
$\pi N$ channel, but puts significant demands on models to interpret
the data consistently.  The objective of this work is to revive the 
multi-channel analysis of partial wave amplitudes into resonance 
parameters of the Carnegie-Mellon Berkeley 
(CMB) group.  Although this is meant to be a significant step toward
a full analysis of all contributing reactions, the equally important 
problem of extracting partial wave
amplitudes from the observables is not considered in this paper.

The baryon resonances (called $N^*$ states from now on in this paper)
extracted from amplitude analyses are thought to be predominantly 
composed of 3 valence quarks because of the SU(3) symmetry seen in the 
spectrum of hadrons of low total angular momentum.  
This notion has been the
basis for developing various quark models, ranging from the well-studied
Constituent Quark Models\cite{isgur,Capstick,iachello,card}, Chiral
Bag Models\cite{brown,thomas}, NJL models\cite{njl}, 
Soliton models\cite{wilets}, to the most recent
Chiral Constituent Quark Models\cite{chiralq}.  All of these 
hadron models are motivated by QCD and are constructed 
phenomenologically to 
describe the hadron spectrum in terms of suitable parameterizations of 
quark confinement mechanisms and quark-quark interactions.
With some additional assumptions on the decay mechanisms, these models can 
use the constructed wave functions to predict the 
decay widths of the $N^*$ states.
For example, the one-body form of quark currents is 
used\cite{Capstick92,iachello} in most of the
constituent quark models in predicting the $N^* \rightarrow \gamma N$ 
decay width.
A $^3P_0$ model is assumed in Ref.\cite{CapstickRoberts} for creation of
$\bar{q}q$ pairs in the calculation of the decay of 
an $N^*$ into nucleons and
mesons. The comparison of these predictions with the 
decay width data, such 
as that listed by PDG, is clearly a more
detailed test of hadron models. 
In recent years, lattice QCD calculations have been extended to predict
masses of the low lying baryons.\cite{Weingarten}. 
Although the lattice QCD calculations are far superior to the 
empirical quark models, 
they are much more difficult to apply to the decay widths.
In any case, precise data of resonance parameters such 
as masses and partial decay
widths for various final states are needed to distinguish QCD-inspired
hadron models and, ultimately, understand the 
non-perturbative aspects of QCD.

The determination of resonance parameters is almost always a two-step
process.  First, phase shift analyses (elastic data) and isobar
analyses (inelastic data) are used to separate the cross 
section and polarization observable data
into partial wave amplitudes, often in the form of $T$ matrices.  
With a large data set, this determination has less 
model dependence than the extraction of resonance properties
in the second step.  If nonresonant effects are small, resonances 
will show up as counterclockwise rotations in the complex energy plane
(Argand diagram) as the energy dependence of 
these $T$ matrices is plotted.   Partial wave amplitudes have been
determined  by various groups for elastic scattering and much less 
often for inelastic scattering.  Older elastic scattering 
analyses of Carnegie-Mellon Berkeley (CMB)\cite{Cutpin} and 
Karlsr\"{u}he Helsinki (KH)\cite{KH80,Koch} stressed the importance of
theoretical constraints such as dispersion relations to ensure the
uniqueness of the fit.  The data situation has improved significantly 
since then; the more recent work of the Virginia Polytechnic Institute 
and State University (commonly called VPI) group is the most visible recent 
effort\cite{VPIpin}.  They have regularly updated the data base and 
attempted to cull
out older less viable results.  They determine single-energy (also
called energy independent) $T$ matrices for partial waves up to L=6 
and satisfy a smaller set of 
dispersion relations than the earlier work.  All these analyses are
available in the VPI repository~\cite{VPIpin}.  We use the VPI 
single-energy elastic partial wave amplitudes from the 1995 analysis 
(SM95) in this work.  At the same time, we recognize the point made by
H\"{o}hler~\cite{hohleraa} that this analysis is not as well constrained as 
the older analyses.  We use the single-energy solutions rather than the
more debatable smoothed solutions.

The second step is to extract the
resonance parameters from the partial wave amplitudes.
In a simple picture, each transition amplitude in a $\pi N$ reaction 
would be parameterized as the product of the excitation
strength of the incident channel to a given resonance and the 
decay strength of the resonance into the allowed final states
with a resonance propagator for the intermediate state.
However, this s-channel resonant mechanism is far from complete since the 
u-channel and t-channel mechanisms, as implied by crossing symmetry or
meson-exchange mechanisms, are known to be important. 
Thus the parameterization of the amplitude in
each partial wave must contain a resonant part and a 
non-resonant part (called the background term in most of the literature). 
Furthermore, the threshold effects associated with each decay 
channel($\pi N, \eta N, \gamma N, \pi\Delta, \rho N, \omega N,
\pi N^*(1440)$ and others) must be treated correctly
within the multi-channel unitarity condition. 
Thus, resonance extraction requires a significant calculational effort and 
many articles have presented various ways to proceed in practice.
The PDG mostly bases its recommended values for 
baryons on a few works that study the full resonance region 
(1170 MeV$<W<$2200 MeV).  These include  
older work by the CMB group\cite{Cut79} and
the KH group\cite{KH80}, 
and more recent work by the Kent State University (KSU) group\cite{ManSal}, 
and the VPI group\cite{VPIpin}.  A very recent work of Feuster and 
Mosel~\cite{Feuster} fits data for $W<$1900 MeV.  
All of these efforts use the data of $\pi N$ reactions.  All maintain 
unitarity, though the methods employed are quite different.  
This is reasonable 
since there is more than one way to implement the unitarity condition.  
However, these 
analyses differ significantly from each other in handling the multichannel 
character of the $\pi N$ reactions.
The CMB and KSU groups use a formalism that allows for many channels,
while the KH and VPI groups  
focus on the $\pi N$ elastic channel.  KH accounts for all inelasticity
in absorption parameters and VPI uses a dummy channel to
account for all inelasticity.  Feuster and
Mosel~\cite{Feuster} fit elastic scattering and  
inelastic scattering cross-section data with asymptotic two-body final states 
directly, and 
account for all the remaining components of the total inelastic
cross section with a dummy inelastic channel.  
For most strongly excited, isolated states, 
these five analyses tend to agree within the assigned errors.
However, as mentioned above, significant differences between them exist
in many cases.  

	This paper will revive the CMB approach\cite{Cut79} and apply it
to extract baryon resonance parameters from partial wave amplitudes 
of $\pi N$ 
reactions with a large variety of final states.  This approach emphasizes
the analytic properties of scattering amplitudes in 
the complex energy-plane 
that are consistent with the dispersion-relation approach and potential
scattering theory.  Other methods of handling multi-channel
unitarity are through the $K$-matrix 
approximation~\cite{VPIpin,Moorhouse,Feuster} and 
the KSU model~\cite{ManSal}. 
Although the transition amplitude is parameterized in a form similar to
these models, there are distinct and important differences.
In particular, the threshold for each of eight possible channels is
treated correctly with 2- and 3-body unitarity requirements imposed.
Thus, resonances can be found as poles in the $T$-matrix, a feature
missing in most $K$-matrix models.
Obviously, the CMB approach is most suitable for
extracting resonances that are close to inelastic channel thresholds.  
Although the KSU approach by Manley and Saleski and more recent 
$K$ matrix models~\cite{Feuster} also account for the structure 
due to channel openings, their multi-channel
parameterization is completely different from the CMB model in realizing
the unitarity condition.  In the CMB model, dispersion relations are
used to guarantee analyticity in the amplitudes.  
The KSU parameterization is an extension 
of a $K$-matrix formulation.  It does not allow the analytic
continuation into the complex energy plane, which is required to find the
resonance pole.  Feuster and Mosel present poles 
from a speed plot analysis, stating that technical problems associated
with their partial wave decomposition prevent 
direct determination of poles from their $K$ matrix analysis.
There is the additional problem that resonances are associated with
poles in the $T$ matrix rather than the $K$ matrix.

In the CMB model, the non-resonant t-channel and u-channel mechanisms 
are simulated by including the transitions to sub-threshold one-particle 
states with masses below the $\pi N$ threshold and with the possibility of 
producing either an attractive or a repulsive potential.  Existing models
treat these effects in a variety of ways.  The empirical method used here
is rather different from the polynomial parameterization 
often used in previous amplitude analyses. 
The consequence of this difference is very significant in practice since 
the intrinsic ``structure" of the non-resonant term due to the channel 
opening is built in correctly in the CMB approach, but can be easily missed
if the ``smoothness" of the background term is the only criterion.  
Modern $K$ matrix methods~\cite{Feuster} fit coupling strengths
in various diagrams.  These have the advantage of fitting fewer
parameters than the more empirical approaches and have the correct partial
wave decomposition of the diagrams include.   The disadvantages include
ignoring all off-shell intermediate state scattering and the inability
to describe resonances of total angular momentum larger than 3/2.

The CMB approach was published in 1979.  Although inelastic data were used
in the analysis, the elastic $\pi N$ scattering data were 
emphasized.  
In this work, we follow closely their theoretical approach, making
only small changes in the parameterization of the amplitudes.
However, we make significant changes in the data set used.
We include all $\pi N \rightarrow \eta N$ data, some of which was
unavailable then.  (They depended on a separate analysis of backward
$\pi^- p$ elastic scattering and other data\cite{Chao} to model the eta cusp. 
This analysis would give different results with the present data set.)  
We also directly fit 
$\pi N\rightarrow \pi \pi N$ inelastic data, represented by 
the amplitudes determined in the isobar model fit of Manley, Arndt, 
Goradia, and 
Teplitz~\cite{Manley84}.  By including 30 percent more
data, the Manley {\it et al.} inelastic amplitudes are more accurate
than those used in the original CMB analysis.  Furthermore, we will use
the most recent VPI energy independent amplitude~\cite{VPIpin} as the 
$\pi N$ elastic scattering input.
This amplitude is significantly different from the one used by the CMB 
group and is also more accurate than the KH amplitude in the 
S$_{11}$ channel, as discussed in Ref.~\cite{hohler94}.  
These differences in the input data make 
our results for some cases (in particular the S$_{11}(1535)$ state) 
significantly different from the CMB values listed in Particle Data Table.

	Batinic {\it et al.}~\cite{Bat} have applied the CMB model to perform 
an analysis with only 3 channels: $\pi N$, $\eta N$, and a dummy channel 
meant to represent the complex set of $\pi \pi N$ channels.  Their focus is 
on the dynamics associated with the $\eta N$
channel.  They fit the KH80~\cite{KH80} energy-independent amplitudes 
of the $\pi N$ elastic data and the $\pi N \rightarrow \eta N$ data.
The $\eta N \rightarrow \eta N$ 
amplitudes are the predictions of the model. 
The Brown {\it et al.} data~\cite{Brown} is the largest body
of  $\pi N \rightarrow \eta N$ data in the energy region close 
to threshold, but it is felt to have
significant systematic errors~\cite{Bat,Clajus}.  Batinic {\it et al.} give it 
a weight 5 times smaller than the other data to deemphasize it.
We will use the same $\pi N \rightarrow \eta N$ data with the same weights 
in our analysis.
This is an important part of the determination of the parameters 
associated with the S$_{11}$(1535) resonance.

	The present work is the first step in an ongoing program to develop
a model appropriate for the new generation of data coming from Jefferson
Lab and other labs.  Although we anticipate further development of the model,  
the goal of this work is to apply the CMB model in a form very similar to 
its original implementation.  We closely follow the procedures of
Ref.~\cite{Cut79}, but use it to determine the baryon spectrum and
decay branching fractions 
from the modern data.  Future publications will address interesting but 
nontrivial issues such as how to determine the ``best'' baryon spectrum and 
how to differentiate between resonance poles, bound state or virtual poles, 
and poles 
that have changed Riemann sheet by moving across an inelastic threshold cut
to the sheet most directly reached from the physical data.

	In section II, we present a detailed account of the CMB multi-channel
unitary model. To illustrate the main features of the model, we
discuss in section III the examples of an 
isolated resonance and the 2 resonance-2 channel situation. 
The full results
are presented and discussed in section IV.  A summary and outlook are
given in section V.  A more complete discussion of the methods and
results is given in Ref. \cite{thesis}.


\section{The CMB Unitary Multi-channel Model}
\label{se:model}

\subsection{Representation of 3-body final states}
The first step of a multi-channel analysis to determine N* 
parameters is to extract from
the $\pi N \rightarrow \pi \pi N$ data a set of
partial-wave amplitudes for the transitions from a $\pi N$ state to
various quasi two-particle channels in which one of the two particles 
is either a $\pi N$ or a $\pi \pi$ resonant state:
$\pi\Delta$, $\rho N$, $\pi N^*(1440)$ and a very broad 
$(\pi \pi)_{J=T=0} N$ channel. 
For convenience, we will use the notation $\sigma$ to label the 
S-wave, isoscalar $\pi \pi$ state.  
This procedure was introduced by the 
Carnegie Mellon-Berkeley (CMB) group in the 1970's.  
A similar procedure 
was used later by the Virginia Polytechnic Institute and State 
University (VPI) and Kent State University 
(KSU) groups\cite{Manley84} to obtain partial wave amplitudes from all of the 
$\pi N \rightarrow \pi \pi N$ data available in 1984. 
These amplitudes, called VPI-KSU amplitudes, will be used in this 
work.  

A more precise analysis should start with 
the original data of $\pi N\rightarrow \pi\pi N$ reactions.
This event data has been stored by the VPI group.  
The raw $\pi N \rightarrow \pi\pi N$ data was not received from
them in time for the present analysis.

\subsection{Model details}
\label{se:moddet}
To extract resonance parameters from VPI-KSU amplitudes, it is
necessary to employ a multi-channel formulation of the $\pi N$ reaction.
This is accomplished in the CMB approach by assuming that the transition 
amplitudes of the $\pi N$ reaction can be 
written in the center of mass (c.m.) frame as
\begin{equation}
	T_{ab}=\sum_{i,j=1}^{N} 
               f_a(s)\sqrt{\rho_a(s)}\gamma_{ai}
                             G_{ij}(s)
               \gamma_{jb}\sqrt{\rho_b(s)}f_b(s)
        \label{eq:tcutkosky}
\end{equation}
where $s$ is the total center of mass energy squared, indices 
$a, b$ denote the asymptotic channels which can be either
a stable two-particle state ($\pi N$, $\eta N$, $K\Lambda$,  $\cdot\cdot$)
or a quasi two-particle state ($\pi\Delta$, $\rho N \cdot\cdot$).
These asymptotic channels are coupled to a set of  
intermediate states (resonances) denoted by indices $i,j$.   
The scattering matrix defined by Eq. (\ref{eq:tcutkosky}) is 
related to the $S$-matrix via $S=1+2iT$ with $S^\dagger S=1$. Hence we have
the following unitarity condition:
\begin{equation}
	\Im m (T_{ab})=\sum_{c=1}^M T^*_{ac}T_{cb}
	\label{eq:Tunitary}
\end{equation}

The crucial step of the CMB model is to 
choose a parameterization of various
quantities in  Eq.~(\ref{eq:tcutkosky}) such that 
Eq.~(\ref{eq:Tunitary}) is satisfied.  
Furthermore, the resulting analytic structure of the 
scattering amplitudes is consistent with the well-developed 
dispersion relations for $\pi N$ elastic scattering and multi-channel
potential scattering theories.  
This is accomplished by using the following prescription. 
First, it is assumed that the $i$th resonance to be found is identified 
with the bare particle, $i$, with a bare mass squared, $s_{0,i}$. 
The strength constant $\gamma_{ai}$ and form factor $f_a(s)$ in 
Eq.~(\ref{eq:tcutkosky}) define the decay of the $i$th resonance
into an asymptotic channel $a$.  The form factor is defined by
\begin{equation}
	f_a(s)=\left(\frac{p_a}{Q_1+\sqrt{p^2_a+Q_2^2}}\right)^{l_a}
	\label{eq:fthresh}
\end{equation}
where $Q_1$ and $Q_2$ are empirical constants defining how quickly a 
channel, with orbital angular momentum $l_a$, opens up\@. 
$p_a$ is the center of mass momentum for channel $a$.
In this analysis, we set $Q_1$ and $Q_2$ equal 
to the pion mass.  These are the same values as were used by CMB;
they chose this parameterization and these values as a result of 
a study of the best way to model the left-hand cut. 

The right-hand or unitarity cut is set through the channel propagators,
$\Im m \phi_a(s)=f_a^2 \rho_a(s)$.  The phase space factor 
$\rho_{a}$ in Eq.~(1) is defined by
\begin{equation}
	\rho_a=\frac{p_{a}}{\sqrt{s}}
      \label{eq:rho}
\end{equation}
for a stable two-particle state. 
For quasi two-particle channels, $\rho_a$ clearly must be defined 
consistently with the phase space factors used in defining
the propagation of this resonant two-body state during the collision, as 
required by the unitary condition Eq.~(2).
In the CMB model, this is achieved by assuming 
that the only interaction during 
collisions is a vertex interaction which converts 
intermediate states into asymptotic states.
Then the propagator $G_{ij}$ in Eq. (\ref{eq:tcutkosky}) 
can be graphically depicted in 
figure~\ref{fig:Dyson} and is defined by the
Dyson equation:
\begin{equation}
	G_{ij} = G^0_{ij} + G^0_{il} \Sigma_{lk} G_{kj}		 
	\label{eq:Dyson}
\end{equation}
Eq. (\ref{eq:Dyson}) is an iterative equation.  A sum over repeated 
indices is implicit.  $G$, $G_0$, and $\Sigma$ all vary with $s$ and are
$N\times N$ matrices, where $N$ is the number of intermediate resonant
and nonresonant) states 
considered.  Each bare intermediate state has a propagator, $G_0$, defined by:
\begin{equation}
	G_{ij}^0(s) = \frac {\delta_{ij} e_i}{s-s_{0,i}}
	\label{eq:bareprop}
\end{equation}
where $e_i=+1$ for states that correspond to the 
resonances that will be fit.  To simulate the $t-$channel
and $u-$channel mechanisms, two subthreshold
bare states with a mass $s_{0,i}$ below the 
$\pi N$ threshold are introduced.
These subthreshold states will simulate an attractive 
background potential
for $e_i=-1$, and a repulsive potential for $e_i=+1$.  In principle, this 
prescription can simulate any $t-$ and $u-$channel mechanisms if a
sufficiently large number of bare states are included.  
This is well known in
potential scattering theory.  The actual number of the subthreshold
states needed is not intrinsically known.  We use two subthreshold 
states and one state at very high energy for every partial
wave.  In the original work, CMB~\cite{Cut79} found negligible 
differences between fits using a (smaller) number of subthreshold states
and fits using actual potentials that simulate the left-hand cut.
They are allowed to couple to $\pi N$ and $\eta N$ asymptotic
states.

The self-energy, $\Sigma_{ij}$, describes the 
dressing of bare particles by the
coupling with two-particle channels, as depicted in 
figure~\ref{fig:Sigdiagram}.  It therefore 
must depend on the strengths($\gamma_{ia}$'s)
and form factors($f_a$'s), and is assumed to 
take the following form
\begin{eqnarray}
	\Sigma_{ij}(s)=\sum_{c=1}^{M} \gamma_{ci}\Phi_c(s)\gamma_{cj}
\end{eqnarray}
For the contributions from stable two-particle channels, we have
\begin{eqnarray}
\Phi_c(s) = \phi_c(s)
\end{eqnarray}
with 
\begin{eqnarray}
\phi_c(s)=
\frac{1}{\pi}\int_{s_{th,c}}^{\infty}dsf^*_c(s')g_c(s';s) f_c(s')
 	\label{eq:selfen}
\end{eqnarray}
Here $s_{th,c}$ is the threshold for channel $c$, and $g_c(s)$
is the propagator of the two-particle channel $c$. 
The task now is to choose
$g_c(s';s)$ such that the unitarity condition Eq.(2) is satisfied and the
desired analytic structure of the amplitude can be generated.  It can be
shown that the $\phi_c(s)$ of the CMB model is obtained by
assuming that
\begin{eqnarray}
g_c(s';s)=\frac{1}{\pi} \left[\frac{\rho_c(s')}{s'-s + i\epsilon} - P
\frac{\rho_c(s')}{s'-s_0} \right]
\end{eqnarray}
where the density of states $\rho_c(s')$ has been 
defined in Eq.(4), and $P$ means taking the 
principle-value part of the propagator.
Substituting Eq.(10) into Eq.(9), we then obtain 
the following dispersion relation for the auxiliary function $\phi_c$
\begin{equation}
	\begin{array}{c}
		\Im m \left( \phi_c(s)\right) =f_c^2(s) \rho_c(s) \\
		\Re e \left( \phi_c(s)\right) = 
                  \Re e \left( \phi_c(s_0) \right) + 
                  \frac{s_-s_{th,c}}{\pi}\int_{s_th}^\infty 
                  \frac{\Im m \phi(s')}{(s'-s)(s'-s_0)}ds'
	\end{array}
	\label{eq:phitrue}
\end{equation}
We see that Eq.(11) is a subtracted dispersion-relation which
has a form similar to what has been established in
many studies of $\pi N$ scattering. 
In the complex-$s$ plane, one can choose the 
subtraction point $s_0$ such that
the resulting scattering amplitude has a pole on the left-hand side 
and a branch cut from $s_{th,{\pi N}}= (m_\pi + m_N)^2$ to $+ \infty$.  
For each channel, we choose the value at threshold for $s_0$, and we set the
subtraction constant, $\Re e \left( \phi(s_0) \right)$, such
that $\Re e \left( \phi_c(s)\right)$ is 0 at threshold.
This arbitrary choice has the primary effect of shifting
the value of the bare mass squared, $s_{i}$, for each resonance
in a partial wave, but does not affect the physical mass of 
the ``dressed'' resonance.
Here we assume that this dynamical assumption is valid for all stable
two-particle channels like $\eta N$, and $K\Lambda$.

For a quasi two-particle channel $c$, the function $\Phi_c(s)$ in Eq.(7) 
must account for the mass distribution of one of the two particles which 
is itself
a resonance state. To be specific, let $m_1$ be the mass of the
stable particle in the channel $c$ and $m_{2}$ and $m_{3}$ be
the masses of the two daughter particles from the decay of the resonant
subsystem into a channel $r$.
Then the form assumed by the CMB model can be more explicitly written as
\begin{equation}
	\Phi_c(s)=\int_{(m_{2}+m_{3})^2}^{(\sqrt{s}-m_1)^2}
                 ds_r \sigma(s_r) \phi_c(s_r)
        \label{eq:phiquasi1}
\end{equation}
where the mass distribution of the quasi particle was taken to be:
\begin{equation}
	\sigma(s)=\frac{\gamma_r \Im m \left( \phi_r(s)\right) / \pi}
               {(M_r^2-s)^2 + \gamma_r^2 \Im m \left( \phi_r(s)\right) ^2}
	\label{eq:weight}
\end{equation}
Here $\phi_r(s)$ is again defined by the subtracted dispersion relation 
Eq.(11) for the appropriate resonant subsystem. Eq.(\ref{eq:weight}) has 
the commonly
used Breit-Wigner resonance form. The coupling strength $\gamma_r$ is
related to the width $\Gamma_r$ of the considered resonance state by
$\Gamma_r = \frac{\gamma_r^2 \Im m \left( \phi(M_r^2) \right)}{M_r}$.  
In this work $r$ is either a $\pi N$
state or a $\pi\pi$ state.  From the empirical
values of the widths for $\Delta$ and $\rho$, mass distribution functions
for the $\pi\Delta$ and $\rho N$ channels can be fixed. For the other
quasi two-particle channels, $\sigma N$ and $\pi N^*(1440)$, the 
width $\Gamma_r$ is fixed at a standard value in the fit\cite{ManSal}.
The above formalism for the quasi two-body channels provides a unitarity
cut along the real $s$-axis from the three-body $\pi \pi N$ threshold 
to infinity.  Furthermore, it makes the resulting scattering 
amplitudes satisfy the unitarity condition Eq.(2). This is how the 
three-body unitarity is implemented in CMB model.

The Dyson equation, Eq.(5), is algebraic and can be solved by
inverting a $N\times N$ matrix.  Schematically, we have
\begin{equation}
	G_{ij}(s) \equiv [H^{-1}(s)]_{ij}
	\label{eq:Gsolve}
\end{equation}
with the matrix element of $H$ defined by
\begin{eqnarray}
     H_{ij}(s) = \frac{s-s_{0,i}}{e_i}\delta_{ij} - \Sigma_{ij}(s)
\label{eq:Heqn}
\end{eqnarray}
Now all of the ingredients needed for calculating 
the $T$-matrix elements of Eq.(1) are in place.  
The variable
parameters (i.e. couplings $\gamma_{ic}$ and poles $s_{0,i}$) are then 
adjusted to fit the VPI-KSU partial wave $T$-matrix elements.

\subsection{Resonance parameter extraction}
\label{se:respar}

Following reference~\cite{Cut79} a resonance position is
identified with a pole of the scattering 
$T$ matrix in the complex energy-plane.  This can only be done
for models that can be evaluated for complex values of $s$, i.e.
models that have the correct analyticity structure.
In the CMB model, the determinant of
the $H$ matrix defined by Eq.(15) equals zero at the
pole position, $s=s_{pole}$, in the complex $s$-plane.
Only the poles located close to the real axis are 
interpreted as resonances.
This procedure involves an analytic continuation 
of $\phi_c(s)$ into the complex $s$-plane for $\Im m (s)<0$.
Clearly, the analyticity, defined by the dispersion relation, Eq.(11), 
plays a ``dynamical" role in finding the resonance parameters. 
This is one of the main differences between 
our approach and the KSU approach. 
 
To proceed, we need to evaluate Eqs.(11) and (12) for complex $s$. 
Each $T$-matrix element has a branch cut beginning at the elastic and 
each inelastic threshold.
The branch cut can have a strong effect on the amplitude, in extreme
cases producing a cusp, e.g. in the $\pi N \rightarrow \pi N$ 
S$_{11}$ partial wave.  Above each threshold, the amplitude is multi-valued.
This is traditionally described by a Riemann sheet structure~\cite{Frazer}.
The amplitude is continuous when analytically continuing to the 
appropriate new sheet
as the value of $s$ crosses the branch cut, but discontinuous when
staying on the same sheet.  The function, $\phi_c(s)$ defined by 
Eqs.(9) and (10) is the channel propagator for the
``first sheet" of the complex $s$-plane, labeled $\phi_{c,I}$.
At $s \ge s_{th,c}$, it has a discontinuity in its imaginary part 
determined by unitarity as $s$ crosses the real axis.  

The resonance pole is on the ``second sheet" in which 
$\phi_c(s)=\phi_{c,II}(s)$; it has the same discontinuity
as the first sheet except for the opposite sign,
\begin{equation}
\phi_{c,II}(s+i\epsilon)- \phi_{c,II}(s-i\epsilon) = \phi_{c,I}(s- i\epsilon)
- \phi_{c,I}(s+ i\epsilon) \ \ \ 
\mbox{for}\, s > s_{th} 
\label{eq:secondsheet}
\end{equation}
It is $\phi_{c,II}(s)$ which is used in the
search of the resonance pole positions.  
Using the same strategy as CMB~\cite{Cut79}, we search the 
$\pi N \rightarrow \pi N$
$T$ matrix for poles on the sheet most directly
reached from the physical region.  As discussed by Cutkosky and
Wang~\cite{Cut90} and elsewhere in the literature, each resonance 
has additional poles on other Riemann 
sheets associated with each inelastic threshold.  However, the 
pole closest to the physical region is most closely associated with the
physical characteristics of the resonance.  The formalism presented here
can be extended to search for poles on other sheets and 
try to distinguish between resonance poles and poles that arise
from bound states of composite particles.  

A resonance pole is found by searching for a zero in the determinant
of the $H$ matrix defined in Eq. (15).  Once a pole is found, $H(s)$ 
is diagonalized at the pole position $s_{pole}$.  This is done to
eliminate resonance-resonance interference effects when multiple resonances
are present in a partial wave.
By using the resulting eigenfunctions $\chi_i$, the
$T$-matrix in the vicinity of the pole can be written as~\cite{Cut79}
\begin{equation} T_{ab}=\frac{B_{ab}-\delta_{ab}}{2i}+
                   \displaystyle{\sum_{ef}}
                       \frac{\sqrt{B_{ae}}\,
                         \sqrt{\Im m (\phi_e)}\,
                           \eta_e\,
                           \eta_f\,
                         \sqrt{\Im m (\phi_f)}\,
                       \sqrt{B_{fb}}}{D(s)}
	\label{eq:TBreit}
\end{equation}
The $\eta_c$ describe the coupling of the resonance to channel c, as 
defined in Eq.~\ref{eq:couplings}.
This form can be shown to be equivalent to the full CMB model.
Eq. (\ref{eq:TBreit}) is a general form for a Breit-Wigner resonance
shifted by nonresonant (background) reaction mechanisms.  To 
determine the resonance properties, we look at
the $T$ matrix in the vicinity of the pole with a simplifying assumption. 
The form of the background part, $B_{ab}$, is 
assumed to be smooth in the immediate vicinity of a pole.  This allows 
us to use
the denominator, $D(s)$, to define the Breit-Wigner resonance parameters 
at the pole, ignoring the background.
The denominator of the above expression is then matched to a relativistic
Breit-Wigner form.  The full denominator can be written as
\begin{equation}
	D(s)=r-s-v\displaystyle{\sum_{c}}y_c\phi_c
	\label{eq:DBreit}
\end{equation}
where  
\begin{equation}
	y_c = \left| \eta_c \right|^2
            = \left| \sum_{i=1}^N \gamma_{ic}\chi_i \right|^2.
	\label{eq:couplings}
\end{equation}
The real constants, $r$ and $v$, are defined by equating $D(s_{pole})=0$.
The resonance mass comes from the real part of the denominator 
and the width comes from the imaginary part.

  The general form of the Breit-Wigner denominator for a multi-channel
situation is
\begin{equation}
    D_{BW} = (W_{res}^2-s) - i \, W_{res} \, \sum_{c} \Gamma_{res,c}
\label{eq:BWdenom}
\end{equation}
where $W_{res}$ is the mass of the resonance and 
$\Gamma_{res,c}$ is the decay width of the resonance to asymptotic state $c$.
Thus, shifts in the resonance mass can be identified with the real parts of
the denominator and the imaginary part 
can be identified with a sum over the partial decay widths of the
resonance to various channels, $c$.  To obtain numerical values, we use 
a linear 
approximation for the real part of the summation in Eq.~\ref{eq:DBreit}.

Using the above definitions, the following qualitative statements
about the relationship between the resonance parameters and
the model parameters can be made.
The resonance width is approximately equal to
twice the imaginary part of the denominator at resonance. 
Because the self energy term in the denominator 
has strong energy dependence,
the physical pole position is shifted from the bare pole by an amount that
depends of the value of the coupling parameter ($\gamma$) and both the 
value and shape of $\Re e \phi(s)$ in the vicinity of the pole.  
The mass of the resonance is further
shifted from the physical pole because the mass 
is determined on the real axis
of the complex energy plane.  
Shifts from the real part of the bare pole to the 
mass can be positive or negative; the size of 
the shift depends on many factors
and can be quite large (more information can be found in 
section~\ref{se:results}).

Eq. (\ref{eq:TBreit}) has an energy dependent pole shift due to
$\Re e\phi(s)$.  We wish to remove this energy dependence to make a 
connection between the full $T$-matrix in Eq. (\ref{eq:TBreit}) and 
a standard relativistic Breit-Wigner shape. 
Making the assumption that the ``real'' part of the term
$\displaystyle{\sum_{c}}y_c\phi_c$
is linear close to the resonance pole:
\begin{equation}
v \Re e \displaystyle{\sum_{c}}y_c\phi_c \approx \alpha + \beta s
\label{eq:rephi_approx}
\end{equation}
the $T$-matrix can be re-written in the form of a relativistic Breit-Wigner
resonance, without an energy dependent pole shift:
\begin{equation} T_{ab}=\frac{B_{ab}-\delta_{ab}}{2i}+
\sum_{ef} \frac{\frac{\sqrt{B_{ae}\, \Im m \phi_e \, \eta_e}}{\sqrt{1+\beta}}
              \, \frac{\sqrt{B_{fb}\, \Im m \phi_f \, \eta_f}}{\sqrt{1+\beta}}}
   {\left( \frac{r-\alpha}{1+\beta}\right) - s - 
   i v\displaystyle{\sum_{c}}y_c \Im m\phi_c/(1+\beta)}
	\label{eq:TBreitalt}
\end{equation}

In Eq. (\ref{eq:TBreitalt}), the quantitative expressions for the 
resonance parameters can then be identified in terms only of quantities 
evaluated at the resonance mass :
\begin{equation}
	\begin{array}{ll}
		\Re e D(M_{res}^2)=0 & 
		\mbox{Defines Resonance Mass $\rightarrow M_{res}$}\\
		\\
		\Gamma=\frac{\Im m D(M_{res}^2)}{M_{res} 
		  \Re e D'(M_{res}^2)}  & 
 		\mbox{Defines Resonance Width} \\
		\\
		\Gamma_c=\frac{y_c \Im m  \phi_c}
		{\sum_a y_a \Im m  \phi_a} \Gamma &
 		\mbox{Defines Branching Fraction into channel c}
	\end{array}
	\label{eq:respars}
\end{equation}
In the above equations, $D'(s)$ is the derivative of $D$ with respect to $s$, 
$\Re e D'(s)=-(1+\beta)$.  
This term accounts for the fact that there is an energy dependence to
$\Re e \phi(s)$, which shifts the pole position.  

This formulation is identical to the generalized Breit-Wigner form that
is the basis of most fitting and theoretical models.  However, we
{\it only} use this form in the immediate vicinity of the pole to determine
the resonance parameters.  The {\it data} are still fit with the full
model and the pole position is then determined from that fit. This type of 
prescription for obtaining the
mass, width and partial widths of a resonance is model dependent.  It
is important to realize that {\it all} definitions of baryon resonance
parameters are model dependent.   
For a highly elastic and isolated resonance, the model is not 
very important.  However, very few resonances fit this
description and we are trying to present a formulation that minimizes the
model dependence.

\subsection{Relationship to other models}
\label{se:othermod}

The CMB model has a number of features that are not included in commonly used
models.  Here, we present a discussion of formulations 
similar to standard models by making approximations to the full CMB 
model.  These simpler formulas will be used in section~\ref{se:modeldep} 
to show the corresponding
model dependence in the extracted resonance properties.

The full CMB model contains a dispersion relation which guarantees the 
analyticity.  The imaginary part of the channel 
propagator function (Eq.~\ref{eq:phitrue}) is the 
relativistic phase space function and the real part is then 
calculated from a dispersion integral.  Models which 
are not analytic~\cite{ManSal} include only the phase space, so  
we set $\Re e \phi(s) =0$ for all $s$ to simulate them.

The full CMB model uses a Dyson equation to allow for conversion 
into open intermediate states (resonant or nonresonant) and open 
asymptotic channels.  The bare propagator (Eq.~\ref{eq:bareprop}) is 
``dressed'' by all the
open intermediate and asymptotic states.  The $K$-matrix 
formulation~\cite{VPIpin,Moorhouse,ManSal,Feuster} uses the
bare propagator of the CMB model in place of the dressed propagator as
a $K$-matrix rather than a $T$-matrix and 
identifies the resonance properties with its parameters.  
A nonanalytic unitary 
multichannel $K$-matrix using a relativistic 
Breit-Wigner form~\cite{Moorhouse} 
can then be constructed for the contribution of resonance $R$ to the 
reaction between initial state $i$ and final state $j$:

\begin{equation}
 K^R_{ij}(s) =  \frac{ \sqrt{\Im m \phi_i(s)} \, \gamma_{R,i} \, 
                     \gamma_{R,j} \, \sqrt{\Im m \phi_j(s)}} {s_R -s}
\end{equation}
where $\Im m \phi_i(s)$ is the product of the form factor and phase space 
for channel $i$ as defined
in Eq. (\ref{eq:phitrue}).  $M_R$ and $\Gamma_{R,i}$ are
the mass and partial width for decay to channel $i$ of resonance $R$; 
the total width is $\Gamma_R$.  
These three physical quantities are defined by
\begin{equation}
\begin{array}{l}
M_R= \sqrt{s_R} \\
\Gamma_{R,i} =   \frac{\Im m \phi_i(s_R) \, (\gamma_{R,i})^2}{M_R} \\
\Gamma_R= \displaystyle{\sum_{i}} \Gamma_{R,i}
\end{array}
\end{equation}
For the nonrelativistic Breit-Wigner case,
\begin{equation}
 K^R_{ij}(s) =  \frac{ \sqrt{\Im m \phi_i(s)} \, \gamma_{R,i} \, 
                 \gamma_{R,j} \, \sqrt{\Im m \phi_j(s)}}
                 {M_R-W}
\end{equation}
where $\Gamma_{R}$ is defined by
\begin{equation}
\Gamma_R= 2 \, \sum_{i} \Im m \phi_i(s_R) (\gamma_{R,i})^2
\end{equation}
In either case, the corresponding $T$-matrix can then be found 
through the standard definition:
\begin{equation}
T=K*(I-iK)^{-1}
\end{equation}
Since both $K$ and $T$ are matrices, there is no simple closed expression
for $T$ corresponding to Eqs. (24) and (26) except in the single channel case.
  
Nonresonant $K$-matrices must 
also be defined.  To maintain unitarity, the full $K$-matrix is obtained 
by adding
all the resonant and nonresonant $K$-matrices (e.g.~\cite{BenMuk95}).
The corresponding $T$-matrix includes effects of resonance interference
and coupling to nonresonant processes, but only on-shell.     

Various theoretical schemes have been built on the $K$ matrix method.
At low energies, the characteristics of the $\Delta$ (P$_{33}$(1232))
and the S$_{11}$(1535) have been determined with an effective Lagrangian 
method\cite{BenMuk95}.  More complete formulations have been 
developed\cite{Sauermann,Feuster}.  In these models, the mesons and
baryons are each fundamental particles.  Although the number of parameters
is reduced from what is required for the more empirical models,
there are still a number of ambiguities in the construction of the
Lagrangian and in the proper development of a multi-resonance,
multi-channel model.

\section{Database}
\label{se:database}

We devote a separate section to the database 
because it plays a critical role
in the results we obtain.  There are many possible 
asymptotic states that can couple to each resonance.  
There are also a number of resonances which couple weakly to the
$\pi N$ channel so that the state is only seen in the inelastic data.
Just as it is important to include various inelastic channels with proper 
threshold effects in the theoretical model, 
it is also important to include as much of
the relevant data as possible with appropriate error bars.

Although it is best to use the original data, 
various partial wave analyses producing $T$ matrix representations of 
the data are available.  This kind of analysis can be done
with much less model dependence than is found in the analysis used for 
determination of baryon resonance properties.  Nevertheless, these 
analyses make choices in the data used in the fits and their absolute
normalizations since not all data sets 
are consistent with each other; these choices can
add error beyond what was in the original data, thus adding uncertainty
to the fit.  On the other hand, fitting partial wave amplitudes allows 
a simpler fitting strategy- 
separate fits can be made for each
partial wave and less computer time is required.  
This is the same procedure chosen by KSU~\cite{ManSal}.  
We choose to fit the single-energy partial wave amplitudes of 
the VPI group for elastic scattering~\cite{VPIpin} and 
the isobar model fits of KSU-VPI for inelastic pion 
production~\cite{Manley84}.  We also make a separate partial wave 
analysis of the $\pi N \rightarrow \eta N$ data.  

\subsection{$\pi N$ elastic data}
Data in this channel is easiest to measure.  Therefore, the data
are more complete and of higher quality than in the other
channels.  In many resonance parameter analyses, these data have a
dominant role in the results.  This takes advantage of using the best
data.  However, the inelastic data must be included to find resonances
that would not be seen in $\pi N$ elastic scattering.  The 
Constituent Quark Model~\cite{isgur,CapstickRoberts} predicts the 
existence of a large number of 
states (roughly a number equal to the number of states seen
to date) and finds that many of them couple weakly to the
$\pi N$ channel.

There have been numerous analyses of the elastic data.  Since 
complete experimental results are not yet available, 
theoretical constraints must be employed to get unambiguous fits.  
Older analyses of CMB~\cite{Cutpin} and KH~\cite{KH80} had strong 
reliance on dispersion
relations to generate unique fits.  KH80 used fixed-t and
fixed-$\theta$ dispersion relations
at many angles.  The real parts of these results were later compared 
with partial-wave
fixed-t dispersion relation predictions by Koch.\cite{Koch}  
Although KH80 results
are quite noisy close to threshold in the high L partial waves, 
there is good qualitative agreement with the additional constraints.
CMB80 uses hyperbolic constraints in the Mandelstam variables.

Although the older works have the best theoretical underpinnings, 
we use the latest $\pi N$ elastic partial wave amplitudes of 
VPI\cite{VPIpin}.  The VPI analysis uses a significantly larger
data set than was available for the earlier 
CMB80 and KH80 elastic analyses.  Most of the new points are at $W<$1600
MeV, but there are also many new results at higher energies\cite{VPIpin}.
Thus, a number of older data points with significantly larger estimated 
errors could be dropped from the fit, resulting in an improved fit.  
These more recent results are consistent with fixed-t   
dispersion relations at $\theta=0$ and low W.  Although improvements
in the VPI analysis are expected, we use this somewhat debatable~\cite{hohleraa} 
approach in this work since this provides a way to include all of the 
data published
since the older work.  Desire for an update of the 1980's work has been 
expressed at conferences for several years, but no such work is in progress.  

\subsection{$\pi N \rightarrow \pi\pi N$ data}

For the best available representation of 
inelastic data, the choices are much more limited than for 
elastic data.  Since there are no model-independent methods for the
3-body final states, isobar models are employed.  Since no recent 
interpretation of the $\pi N \rightarrow \pi \pi N$ data is presently 
available, we use the quasi-two-body channel decomposition of
Manley, Arndt, Goradia, and Teplitz~\cite{Manley84}.  
That work fit an isobar model to the 
$\pi N \rightarrow \pi \pi N$ data, isolating 
the contributions of $\pi \Delta$, $\rho N$, $\sigma N$ (with $\sigma$ 
representing the $\pi\pi$ strength in an isoscalar $s$ wave), and 
$\pi N^*(1440)$ channels.  Although the $\sigma N$ channel will absorb some
of the nonresonant $\pi\pi N$ strength, these choices ignore some of 
the non-resonant 
$\pi \pi N$ strength and states such as $\pi N^*(1520)$ 
that might be expected to share strength.  They used a data sample of about 
241,000 events spread over 1320-1910 MeV in W; statistical accuracy is much
poorer than for the elastic channel and there is no data at the highest
energies where resonances are found.  All data published after this 
analysis were very close to threshold.  It 
will be clear from our analysis that these data need augmentation and 
that the isobar analysis should be repeated.  

As with the choice in elastic data amplitude, this choice is a compromise
between including all the existing data in a simple form and using a
more proper analysis.  An analysis the $\pi N$ inelastic data
without the assumption of the isobar
model would be extremely difficult and unjustified with the quality of
the present data set.  

In the CMB analysis~\cite{Cut79,Cut90}, the $\pi N \rightarrow \pi \pi N$ 
data were weighted by a factor of 1/3 smaller than the results of 
Manley, {\it et al.}~\cite{Manley84} because they felt the
errors were understated.  We have a similar attitude and weight the
inelastic data by a factor of 1/2.  Otherwise, the fits to the higher 
quality elastic data are degraded.

\subsection{Description of the $\pi N \rightarrow \eta N$ Analysis}

The most significant inelastic channel at low W is the 
$\eta N$ state; data involving this channel is 
thought to be crucial in analyzing S$_{11}$(1535) because the s-wave 
$\pi N \rightarrow \eta N$
cross section is large and rapidly changing close
to the resonance mass.  Unfortunately, the
data in this channel is both limited and of 
uncertain quality.~\cite{Clajus}  The data 
set with the most points close to threshold was 
published by Brown, {\it et al.}~\cite{Brown}.
Clajus and Nefkens~\cite{Clajus} argue that these 
data have unknown errors in
the assigned values of W, making them unusable.  
As mentioned earlier, Batinic {\it et 
al.}~\cite{Bat} put a very small weight on these 
data points in their partial wave fit to the
$\pi N \rightarrow \eta N$ data.  To produce partial wave amplitudes 
for this reaction, we reproduce 
the Batinic {\it et al.} fit to the $\pi N \rightarrow \eta N$ 
data, effectively leaving out the controversial 
Brown {\it et al.}~\cite{Brown} data set.  

Since there is very little data available for the reaction 
$\pi N \rightarrow \eta N$, we use a simplified version of the
full CMB model to simultaneously fit the $\pi N$ elastic T
matrices and the $\pi N \rightarrow \eta N$ data to provide
a parameterization for each of the partial waves with $L \leq$4 
contributing to the $\eta$ production cross section.
The partial wave analysis followed
the procedures used by reference~\cite{Bat}, but used the newest 
VPI $\pi N$ elastic partial wave amplitudes~\cite{VPIpin} instead 
of the older Karlsr\"{u}he Helsinki (KH80) amplitudes.
The channels used in the analysis are $\pi N$, $\eta N$, and a dummy channel
consisting of a fictitious meson with mass chosen so that the dummy
channel opens at about the energy where $\pi N$ inelasticity due to channels
such as $\pi \pi N$ start.  The mass of that fictitious meson changes
from partial wave to partial wave.  The values 
used in the analysis are given in Ref.~\cite{Bat}.
The partial waves used are the I=1/2 partial waves through G$_{17}$.
All partial wave parameters were varied simultaneously.  
More details on this procedure given in this 
section can be found in the references~\cite{Bat,thesis}.

 For this process, the S$_{11}(1535)$ resonance
makes up most of the total cross section.  Therefore, the S$_{11}$ partial
wave is the most accurately extracted.  The other partial wave amplitudes 
are smaller and less accurately determined.  The results of this 
fit are $T$-matrices that
best model the partial wave data.  The best fit is shown as 
an error band in the figures of the results section.  For the final
fits, 40 data points between threshold and 2.3 GeV were used for each 
partial wave.  This insures that these
data provide the appropriate contribution to the total chisquare. 


\section{Illustrative Examples- P$_{33}$ and S$_{11}$ Partial Waves}
\label{se:s11}
	The purpose of this section is to 
introduce features of the model through
the examination of specific partial waves.
The P$_{33}$ and S$_{11}$ partial waves are chosen for this purpose.  
The  P$_{33}(1232)$ or $\Delta$ is the 
best example of an isolated and elastic, i.e. simple, resonance; its 
characteristics are largely well-established.  On the other hand, the 
S$_{11}$ partial wave has some of the most interesting
structure of all the partial waves contributing to $\pi N$ scattering.
Because of this structure, extra care must be taken when extracting 
resonance parameters for this partial wave.  Some of the interesting
features exhibited in this partial wave are 
listed below.
\begin{enumerate}
	\item There are 2 PDG 4* resonances (S$_{11}(1535)$
		and S$_{11}(1650)$) which overlap significantly.
	\item The S$_{11}(1535)$ has strong coupling to both
		the $\pi N$ and $\eta N$ channels and is very
		near the $\eta N$ threshold ($\approx$1487 MeV).
		This produces a strong cusp in the $\pi N$ elastic 
		S$_{11}$ $T$-matrix element.
	\item There are 7 decay channels that have measurable
		coupling to the S$_{11}$ states, each of which
		has different phase space which can cause
		structure in all of the other channels via 
		unitarity.
\end{enumerate}
All of the above features are adequately handled in the CMB Model.

	This section will help in understanding the 
unitarity, analyticity, and other properties of the model, which
are especially important in the S$_{11}$ partial wave.
A discussion of the cusp associated with the S$_{11}$
partial wave will also be given.  Furthermore, this section will 
present a systematic study of the model dependence of 
resonance parameters  extraction by examining results obtained
when leaving out various features of the full CMB model.

\subsection{Single Channel-Single Resonance Case}
\label{se:singres}

	The equations from section \ref{se:model} give
expressions for a reaction $T$-matrix element for any number of open
scattering channels or asymptotic states ($\pi N$, $\eta N$, etc.) 
and any number of intermediate states (S$_{11}(1535)$, 
P$_{33}(1232)$, etc.).  The $T$-matrix
elements for one asymptotic state and one
intermediate state are much simpler.  The resulting equations can then 
be applied with good success to the 
P$_{33}$ partial wave near the $\Delta(1232)$ resonance.
Despite the complexity of the CMB formulation, the $T$-matrices for an
isolated, single channel resonance have a form similar to that
found in the Breit-Wigner shapes commonly used.  

	The one asymptotic state-one intermediate state case requires
two parameters (using $\Delta$ to label
the P$_{33}(1232)$ intermediate state and $\pi N$ as the 
asymptotic state), one coupling $\gamma_{\Delta,\pi N}$
and one {\it bare} pole energy, $s_{\Delta}$.
Since there is only one intermediate state the matrix equations
are reduced to scalar equations and the math is simple.
There is only one 
channel propagator $\phi_{\pi N}$ which is determined analytically
with a contour integral.  
(There is a functional form for the channel propagator
only in the S-wave case.  The results for the $\pi N$ channel in an
S-wave are shown in figure~\ref{fig:phi}.)  There is also one
self energy term according to Eq.~\ref{eq:selfen}.
\begin{equation}
	\Sigma_{\Delta,\Delta}(s)=
	\gamma_{\Delta,\pi N}^2 \phi_{\pi N}(s)
\end{equation}
Since the function $\phi$ varies with energy, 
$\Sigma$ does also.
The $H$ matrix is then defined by:
\begin{equation}
   \label{eq:hmatrix}
   \begin{array}{rl}
     H= & (s_{\Delta}-s)-\Sigma_{\Delta,\Delta} \\
      = & (s_{\Delta}-s)-\gamma_{\Delta,\pi N}^2 \phi_{\pi N}
   \end{array}
\end{equation}
The $G$ matrix (the dressed propagator) is then calculated from 
$H$ as defined by 
 Eqs. (\ref{eq:Gsolve}) and (\ref{eq:Heqn}) in section \ref{se:model}:
\begin{equation}
   G=\frac{1}
   {(s_{\Delta}-s)-\gamma_{\Delta,\pi N}^2 \phi_{\pi N}}
\end{equation}
$T_{\pi N,\pi N}$ is defined using Eq. (\ref{eq:tcutkosky})
and the above equations:
\begin{equation}
   T_{\pi N,\pi N}=\frac{\gamma_{\Delta,\pi N}^2 \Im m \phi_{\pi N}}
     {(s_{\Delta}-s)-\gamma_{\Delta,\pi N}^2 \phi_{\pi N}}
\label{eq:teqn}
\end{equation}

The results of a model calculation for the $T$ matrix is shown in 
Fig.~\ref{fig:p33}
with a solid dot at the physical resonance mass.
The equations produce the characteristic shape of a resonance 
with a maximum in the imaginary part and a 
zero in the real part at the resonance mass.  
A similar signal should 
be seen for strong resonances, but there will be shifts 
in the real data when underlying backgrounds of varying smoothness 
are included.  

Eq.\ref{eq:teqn} has the usual form of a relativistic
Breit-Wigner resonance as expected, but there are 
some important new features.
The real part of the pole is shifted 
from the bare pole energy ($s_\Delta$) 
by $\gamma^2 \Re e \phi(s)$ and
the pole gains an imaginary part, $\gamma^2 \Im m \phi(s)$ due
to couplings of the resonance to the asymptotic state as it propagates.  
In this
model, these shifts are energy dependent and come from both unitarity 
and analyticity requirements.  The analyticity condition, Eq.~(11)
then allows an analytical continuation of the above expressions 
to the complex $s$ plane where the
{\it physical} pole position, mass, and width of the 
resonance must then be 
determined by a search discussed in section~\ref{se:model}.  

In other contexts, unitarity requirements 
are satisfied through 
inclusion of final state interactions.  These methods add terms to the 
denominator similar to what comes from the CMB model.  However, these 
models are not analytic.  

Near the resonance peak,
Eq.~\ref{eq:teqn} can be expressed in terms of a mass and width
identical to a generalized Breit-Wigner shape.  Thus, the features
of the CMB model can be absorbed into effective constants for the
case of an isolated resonance.

\subsection{Two Channel- Two Resonance Case}
\label{se:twochtwores}
	The equations get rapidly more complicated as the number of 
resonances and open channels in a partial wave increases.  The
two channel (or asymptotic states) 
two resonance (or intermediate states) case is still instructive.
The relevant equations are given below for the $L=0$ isospin 1/2 (S$_{11}$)
partial wave.  In reality, this partial wave has two strong channels and
two strong resonances.  The results shown below 
use realistic
parameters for these states, but are not meant to be an accurate
representation of data because non-resonant processes and other
channels are ignored.  For 
clarity, channel labels $\pi N$ and $\eta N$ and 
resonance labels of 1535 (referring to S$_{11}(1535)$) 
and 1650 (referring to S$_{11}(1650)$) are used.
Parameters that must be determined include four coupling strengths  
($\gamma_{1535,\pi N}$, $\gamma_{1535,\eta N}$,
 $\gamma_{1650,\pi N}$, and $\gamma_{1650,\eta N}$)
and two bare poles (S$_{1535}$ and S$_{1650}$).

The terms in the $\Sigma$ self energy matrix are constructed
from Eq. (7):
\begin{equation}
   \begin{array}{rl}
      \Sigma_{1535,1535} & 
           =\gamma_{1535,\pi N}^2\phi_{\pi N} +
            \gamma_{1535,\eta N}^2\phi_{\eta N} \\
      \Sigma_{1535,1650} & 
           =\gamma_{1535,\pi N}\gamma_{1650,\pi N}
                               \phi_{\pi N} +
            \gamma_{1535,\eta N}\gamma_{1650,\eta N}
                            \phi_{\eta N} \\
      \Sigma_{1650,1650} & 
           =\gamma_{1650,\pi N}^2\phi_{\pi N} +
            \gamma_{1650,\eta N}^2\phi_{\eta N} \\
   \end{array}
\end{equation}

The $\Sigma$ matrix is symmetric by construction so it takes on
the simple form:
\begin{equation}
 \Sigma (s)=
   \left(
      \begin{array}{cc}
         \Sigma_{1535,1535} &
         \Sigma_{1535,1650} \\
         \Sigma_{1535,1650} &
         \Sigma_{1650,1650} 
      \end{array}
    \right)
\end{equation}
Fig.~\ref{fig:S11sigma} shows the $\Sigma_{1535,1535}(s)$ function.
It is a weighted sum of $\phi_{\pi N}$ and $\phi_{\eta N}$, 
shown in Fig.~\ref{fig:phi}, and thus carries the 
threshold behavior of both the $\eta N$ and $\pi N$ channels.
The other elements have similar qualitative behavior,
but varying weights of $\phi$ because the two resonances couple to the 
two asymptotic channels with different strength. 

The $H$ matrix becomes:
\begin{equation}
 \begin{array}{rl}
   H(s)= & (s_i-s)-\Sigma(s) \\
    = &
   \left(
      \begin{array}{cc}
         s_{1535}-s-\Sigma_{1535,1535} &
         - \Sigma_{1535,1650} \\
         - \Sigma_{1535,1650} &
         s_{1650}-s-\Sigma_{1650,1650}
      \end{array}
   \right)
 \end{array}
\end{equation}
The dressed propagator, the $G$ matrix, has a simple form since
it is a $2\times 2$ matrix:
\begin{equation}
   \begin{array}{rl}
	G= & H^{-1} \\
         = & \frac{1}{\left| H \right|}
	\left(
        \begin{array}{cc}
           s_{1650}-s-\Sigma_{1650,1650} &
            \Sigma_{1535,1650} \\
            \Sigma_{1535,1650} &
           s_{1535}-s-\Sigma_{1535,1535}
	\end{array}
        \right)
  \end{array}
\end{equation}
where $\left| H \right|$ is the determinant of $H$:
\begin{equation}
     \begin{array}{rl}
	 \left| H \right|  = &
          \left(
          s_{1535}-s-\Sigma_{1535,1535}
          \right)
          \left(
          s_{1650}-s-\Sigma_{1650,1650}
          \right) \\
        &  - \Sigma_{1535,1650}^2
     \end{array}
\end{equation}
The $T$ matrix elements are then defined in
section~\ref{se:model} in Eq. (\ref{eq:tcutkosky}).
Explicitly the $\pi N$ elastic $T$-matrix element is:
\begin{equation}
\large
   \begin{array}{rl}
      T_{\pi N,\pi N} = & \frac{\gamma_{\pi N,1535}^2 \Im m \phi_{\pi N}}
         {s_{1535}-s-\Sigma_{1535,1535}- \frac{\Sigma_{1535,1650}^2}
                 {s_{1650}-s-\Sigma_{1650,1650}}} + \\
      & \frac{\gamma_{\pi N,1650}^2 \Im m \phi_{\pi N}}
         {s_{1650}-s-\Sigma_{1650,1650}- \frac{\Sigma_{1535,1650}^2}
                 {s_{1535}-s-\Sigma_{1535,1535}}} + \\
      & \frac{2 \gamma_{\pi N,1535} \gamma_{\pi N,1650} \Sigma_{1535,1650} 
                 \Im m \phi_{\pi N}}
                 {(s_{1535}-s-\Sigma_{1535,1535})
                 (s_{1650}-s-\Sigma_{1650,1650})- \Sigma_{1535,1650}^2}\\
   \end{array}
   \label{eq:T2by2}
\normalsize
\end{equation}
Note that all the functions used in this expression depend on $s$.
The above expression is the sum of two resonances with
an interference term.  (If there is only one resonance, this formula
simplifies to the same as Eq. (32).)
The terms that look like single resonances are
more complicated than in the previous example.  
The self energy terms have the
necessary analytic cuts from the $\eta N$ as well as the
$\pi N$ channels and also make the amplitude unitary.  There
are analogous $T$-matrix elements for the processes: 
$\pi N \rightarrow \eta N$ and $\eta N \rightarrow \eta N$.

In this case, two poles must be found in the complex energy plane.
Since the two resonances have 
significant interference, the non-diagonal elements of the
$G$ matrix are large.  This means the second resonance 
has a contribution to the propagator of the first resonance at the pole
of the first resonance.  Referring to Eq. (\ref{eq:T2by2}), there are
shifts to the mass and width of each resonance due to the presence of
the other in addition to those due to the asymptotic channel couplings
which were not seen in the one channel-one resonance case.
Here, the $H$ matrix (the inverse of $G$ or
the ``denominator'' matrix) must then be diagonalized at each pole to 
isolate the contributions from each individual resonance. 

The $\pi N$ elastic and $\pi N \rightarrow \eta N$ $T$ matrices in
this partial wave corresponding to the above equation
for representative parameters (see Table \ref{tb:s112by2m} and
\ref{tb:s112by2br} for a complete listing of the relevant values)
are shown in Fig.~\ref{fig:s11model}.
The eta threshold has a significant effect on the  
observables.  Therefore,  the
$\pi N$ elastic $T$-matrix element peaks at the $\eta N$ threshold rather
than at the peak of the S$_{11}(1535)$ resonance.
The physical masses are shown as solid dots which are near the peak
of the cross section and the imaginary part of $T$ for the higher state,
but above those positions for the lower state.  
For all channels, the peak
of the $T$-matrix is shifted by a few dozen MeV from the bare mass
by the self energy terms.  (The S$_{11}(1535)$ resonance shape is 
also modified by
the presence of a strong inelastic threshold.) 
There is also significant interference between the two resonances.
Both cross section bumps are unusually
narrow compared to the physical widths because of interference effects.    
We find that the S$_{11}(1535)$ properties are significantly altered
by the interference with S$_{11}(1650)$, similar to the findings
of Sauermann, Friman, and N\"{o}renberg~\cite{Sauermann} 
in a $K$-matrix photoproduction calculation.

\subsection{Cusp structure in the S$_{11}$ partial wave}
\label{se:cusp}
There are two main causes of the strong cusp structure
observed in $\pi N$ elastic scattering differential cross
section, as well as the S$_{11}$ partial wave amplitudes.  
The first is  that the S$_{11}(1535)$ resonance couples
strongly to both the $\pi N$ and $\eta N$ channels and
the $\eta N$ channel threshold ($\approx$ 1487 MeV) is just below 
the 1535 pole.  The second is that the
orbital angular momentum is zero ($\pi N$ S-wave); thus,
the cross section in this partial wave increases
linearly with momentum.  Therefore, through analyticity and
unitarity, a cusp structure 
appears in the $\pi N$ elastic channel.  

Figure~\ref{fig:S11sigma} shows the self energy term labeled
$\Sigma_{1535,1535}$ for the 2 channel 2 resonance model 
of Sec.~\ref{se:twochtwores}.  
This self energy has the appropriate analytic phase space
factors and therefore shows a cusp.  Since $\Sigma$
enters directly into the $T$-matrix elements, the
cusp shows up there as well.  Figure~\ref{fig:S11rbf} shows the
$T$-matrix elements for elastic scattering and $\rho$ production.  
Although the cusp structure is apparent in the $T$ matrices for all 
channels other than that of eta production, 
it is most evident in 
the $\pi N$ elastic channel because most of the 
decay width of the S$_{11}(1535)$ is split roughly equally between
the $\pi N$ and $\eta N$ channels.  All other inelastic channel
openings have the potential of creating a cusp, but this is
by far the strongest case.  At present, the $\pi N$ data is not of high
enough quality to see any other cusp structures.


\subsection{Model Dependence in Analysis of the S$_{11}$ partial wave}
\label{se:modeldep}

	The variation between this and other models can best be seen in an
analysis of the data in the $\pi N$ S$_{11}$ channel, as discussed in
previous sections.
Because of the peculiarities of this channel, data for both 
$\pi N \rightarrow \pi N$ and $\pi N \rightarrow \eta N$ reactions
would be required for a high quality determination of the S$_{11}(1535)$
parameters and data for 
$\pi N \rightarrow \pi \pi N$ would be required for 
a good determination of properties for the higher energy states.  
Presently, the inelastic data 
is of much less quality than for the elastic channel.  

	To test model dependence in the resonance parameters, we have 
analyzed all available and subsets of the data for the S$_{11}$ channel with
the full CMB model and various approximations that simulate the models
employed by various other groups.  Although the final
fits were of similar quality for all cases, there can be large
differences in the extracted resonance parameters.   

	In Tables~\ref{tb:s11mod1} and \ref{tb:s11mod2}, masses,
widths and branching fractions are given for the two of the three S$_{11}$
resonances used in this analysis for a number of different
model types using various data sets to constrain the fits.  
The four columns on the left of the table 
describe features of the model used 
and what data was used in the fit.  The five columns on the right give
the results of the fit for the resonance parameters- mass, width, and
branching fractions.

	The {\it Unitarity} column labels how unitarity
was imposed in the fits.  {\it $K$-matrix} means that unitarity was
imposed by commonly used $K$-matrix methods of Moorhouse, Rosenfeld, and
Oberlack~\cite{Moorhouse}.  For the results shown in the tables, we
recreate the $K$-matrix fit of Ref.~\cite{Moorhouse} 
with the appropriate phase space factors ($\phi(s)$) from our model
as discussed in Sec.~\ref{se:othermod}.
{\it Dyson equation} means
that unitarity was imposed using the Dyson-equation approach
of the CMB model.  Channels can interact any
number of times in various forms before finally decaying 
because the Dyson equation is iterative.

	The column labeled {\it Disp Rel} refers to whether
or not the dispersion relation was used to make
the phase space factors in the self-energy terms 
analytic functions or not.  If the dispersion
relation is evaluated, the phase space factors and hence
the self-energies are analytic functions of the square of the CM 
energy, $s$; otherwise they are not.  When using the
Dyson equation and not evaluating the dispersion relation 
(in the fits, we set $\Re e \phi$ = 0.0), all intermediate 
interactions in the scattering process occur on-shell.
Note that this does not affect unitarity.  If unitarity is achieved
through $K$-matrix methods, the mass and widths are direct parameters
in the fit.  

	The column {\it Res Type} describes the
form of resonance used.  {\it NRBW} refers to a
non-relativistic Breit-Wigner shape and {\it RBW} means a 
relativistic Breit-Wigner shape was used.  For the NRBW case, 
a resonance has a $\frac{\gamma^2}{w_0-w-i\gamma^2}$
form and the $\gamma$ couplings have units of $\sqrt{energy}$.
For the RBW case, a resonance has a 
$\frac{\gamma^2}{s_0-s-i\gamma^2}$ form; the $\gamma$ 
couplings have units of energy.

	Finally, the column labeled {\it Channels in fit}
describes the types of data used in the fit.  For the $\pi N$ case,
only the VPI $\pi N$ elastic S$_{11}$ $T$-matrix elements 
were used in the fit.  For the $\pi N$, $\eta N$ cases
both the VPI elastic data as well as constraints from
a partial wave analysis of $\pi N \rightarrow \eta N$
done by this group were used.  For cases labeled {\it All},
$\pi N \rightarrow \left\{\pi N, \eta N, \pi\pi N \right\}$ are
all included in the fit.  The $\pi\pi N$ channel is composed
of quasi two body channels $(\rho_1 N)_{S}$,
$(\rho_3 N)_{D}$, $(\pi\Delta)_{D}$, 
$\sigma N$, and $\pi N^*(1440)$.  Partial
wave quasi-two-body $T$-matrix elements of Manley 
{\it et al.}~\cite{Manley84} are used for these channels.

	The last row in each table contains results for
our full model.  All channels are used in the fit, unitarity comes from 
use of the Dyson equation, 
analytic phase space factors are used, and
the bare resonance has a {\it RBW} dependence.

	Significant model dependence is seen along with sensitivity
to the data used.  
The lower state (S$_{11}(1535)$) has significant 
model dependence as noted above while S$_{11}(1650)$ tends to have
poorer quality data, i.e. missing data or data 
with error bars that are too small.
We do not show any results for the third resonance 
(S$_{11}(2090)$) because 
the data quality dominates the fitting, causing 
wide variation in fit results, 
e.g. masses vary between 1509 and 2028 MeV.
The data quality for this partial wave is discussed in detail in 
section~\ref{se:detailedresults}.  

We note that without the dispersion relation, the $T$ matrices for 
the $K$-matrix model and the
model using the Dyson equation for the resonance propagator are
equivalent.  Therefore, only the $K$-matrix results without the dispersion
relation are given in the table.  
In general, the choice of relativistic vs. 
nonrelativistic shape for
the bare Breit-Wigner resonance does not have a strong influence for an
isolated resonance such as the 1535 MeV state.  However, the 1650 MeV 
state has a weaker signal (in part because of poorer data quality) and the
two shapes can produce larger differences.  Even there, the agreement in the
case where all data is used (line 3 for NRBW vs. line 6 for RBW) is very
good.

More important differences are found when comparing $K$-matrix vs.
Dyson equation results with the dispersion 
relation included (e.g. line 6 vs.
line 9).  The former is close to the model employed by Manley and 
Saleski~\cite{ManSal}.   These two models have differences of about 
10\% in the total width and up to 50\% in the branching fractions.

The most important deviation from the full result comes from the use
of a truncated data set.  For the 1535 MeV state, ignoring the interference
with the $\eta N$ final state causes the model to fit the Breit-Wigner
shape to the cusp at the $\eta N$ threshold.  The VPI work~\cite{VPIpin} 
has a very small width for the 1535 MeV state; although the $\eta N$ 
channel is mocked up, none of the actual data is used.  For 
even the full model, leaving out the $\pi \pi N$ final state data 
(such as was done by Batinic {\it et al.}\cite{Bat}) produces 20\% 
deviations in the branching fractions.  
We reproduce the updated results of the Batinic {\it et al.} 
paper\cite{Bat}.

Both the mass and the width of the S$_{11}(1535)$
tend to increase as more data channels are added into
the fit.  It is interesting to note that the small
widths (i.e. $<$ 100 MeV) are in situations where only the
$\pi N$ elastic data is used in the fit.  
Also, the branching fraction of $\eta N$ tends to
be smaller than that of $\pi N$ unless
the analyticity is taken into account in the
phase space factor.  In other words, when the 
cusp is handled appropriately, the $\eta N$ channel
becomes the dominant decay mode.  The Manley-Saleski~\cite{ManSal}
analysis, which does not address the analyticity issue, 
is unable to match the cusp well and
concludes that $\pi N$ is the dominant decay mode.

	The S$_{11}(1650)$ and the third S$_{11}$
resonance have more random shifts in their 
resonance parameters.  At energies in the region of the
excitation of the third S$_{11}$ state, only elastic
and $\eta$ production data exist and the signal is not strong.  
Therefore, the fit parameters for the third S$_{11}$ state
are largely determined by 
fitting non-statistical fluctuations of the data.  The fitted masses
vary widely from case to case and we feel the results 
with the different models do not give information about features 
of the models.

\subsection{Elastic Data Dependence}

A major question is the balance between model dependence and data
dependence.  The previous section showed the model dependence using
various subsets of the data used in this analysis.
In this section, we present results for fits of the S$_{11}$ channel
using two different sets of elastic partial wave amplitudes.
In Table~\ref{tb:elas_dep}, we compare our standard results with the 
results of a fit using the combination of CMB80 and KH80 
data sets.  This fit uses the same elastic data 
as Manley and Saleski~\cite{ManSal} except for the additional
$\pi N \rightarrow \eta N$ data.  The differences seen in the 
table are somewhat similar (in total width) to but somewhat smaller 
(in branching fraction) than those presented in the previous section, 
verifying that the major difference between the present work and
Manley and Saleski is likely due to model dependence.  The changes in 
the results
are larger for the second state than for the lowest state, but
the differences are significant for both states.  The differences
are much larger for the third S$_{11}$, but are not shown because
of the weak evidence for this state.

The balance between the branching fraction into $\pi N$ and $\eta N$
for the lower state are very similar in the two cases, but the
total width found using the older data is 22\% larger.  This
gives more evidence for the findings of the previous section that
differences in branching fractions are a primary result of the 
difference between this model and that used by Manley and Saleski.

\subsection{The nonresonant amplitude}

	A major problem in any extraction of resonance parameters is in
the careful separation of resonant and nonresonant mechanisms.
Ideally, an independent calculation would be used for the nonresonant
diagrams.  Here, we choose a more objective strategy and use
a smooth background that can then be ``dressed'' to contain the
correct threshold behavior.  To do this, we add two subthreshold
resonances (one repulsive and one attractive) and one very high energy 
resonance in each partial wave
as `bare' propagators.  These propagators are then dressed identically to
the true resonances.  More details are given in section~\ref{se:moddet}.  

	The separation into resonant and nonresonant components is
shown in Fig.~\ref{fig:S11rbf} for various reactions in the S$_{11}$
partial wave.  We show the magnitude of the $T$-matrix for four
different final states calculated with final fit parameters.  The
three lines shown correspond to including only nonresonant couplings,
only resonant couplings, and all couplings.  Since the nonresonant
and resonant processes are intermixed in the Dyson equation, there
is no way to sum them to get the full result.  

	While some reactions are dominated by resonant processes (e.g.
elastic scattering), others are dominated by the nonresonant 
processes (e.g. $\pi N \rightarrow \rho_1 N$).  The resonance 
excitation must be sampled through a variety of channels 
to provide the full picture.  At very high energies (W$\sim$1.9 GeV),
the lack of data allows the nonresonant processes to dominate.
Resonance extraction at these high energies is very 
unclear with the data presently available.


\section{Results and Discussion}
\label{se:results}

We have applied the CMB model to the database presented in section 
\ref{se:database}- $\pi N$ single-energy elastic $T$ matrices of 
VPI\cite{VPIpin}, the inelastic $T$ matrices of 
Manley, Arndt, Goradia, and  Teplitz\cite{Manley84}, 
and our own partial wave analysis of the $\pi N \rightarrow \eta N$ data.  
The $\pi N \rightarrow \pi \pi N$ 
raw data was not available in time for the present analysis.  A
reanalysis of the $\pi N \rightarrow\pi \pi N$ data is in 
progress~\cite{MuellerMulhearn} by our group; a more 
complete analysis can be presented when that work is finished.
   
The analysis presented here contains features of CMB\cite{Cut79} and
KSU\cite{ManSal} since we use the formalism of the former and a 
data set similar to that used by the latter.  However, the present
analysis goes beyond any previously published.
We present general results and
a detailed discussion of the 
D$_{15}$, D$_{13}$, and S$_{11}$ partial waves.
D$_{15}$ is an excellent example of an isolated resonance, but has strong 
inelastic couplings.  
The D$_{13}$ and S$_{11}$ partial waves each have a strong state 
(well understood for the former and poorly understood for the latter) 
along with less well understood states; the interpretation of 
most of these states are sensitive to the features of the
present model.

We will compare our results to previous work with an emphasis on works
that treat many channels.  We also compare to the composite results 
of PDG~\cite{PDG}.

\subsection{Details of Fitting}
Although the $\pi N \rightarrow \pi \pi N$ data set used here is identical
to that used by KSU, they do not include the 
$\pi N \rightarrow \eta N$ data and
use older elastic analyses.  KSU includes an $\eta N$ channel in the 
S$_{11}$ partial wave fit, 
using the requirements of unitarity to fix its coupling
strength to the resonances.  
For the elastic channel, they use CMB and KH80 
elastic amplitudes simultaneously.
 
For the inelastic channels involving two pions, we define channels  
identical to those chosen in Ref.~\cite{ManSal}.  They are listed in 
Table~\ref{tb:chans}.  Each channel has a specific orbital and spin 
angular momentum; the nomenclature used is given in the table.
These $\pi N \rightarrow \pi \pi N$ 
channels are almost identical to those
chosen in Ref.~\cite{Manley84}.  
The only difference is in the choice of the mass
and width of the fictitious isoscalar-spin 0 meson, 
taken as 1 GeV for each parameter in
Ref.~\cite{Manley84}; like Ref.~\cite{ManSal}, 
we choose a mass and width of 800 MeV.  
We agree with their conclusion that the results are not
sensitive to this choice.

The number of states 
sought in each partial wave was the same as used by KSU\cite{ManSal}.
We also included the same number of open channels in
each partial as KSU because of
the choices in the amplitudes fitted in Ref.~\cite{Manley84}.  
If no amplitudes are provided by Ref.~\cite{Manley84} for a
partial wave, e.g. for F$_{17}$, an appropriate 
dummy channel is used to absorb the flux.  Table~\ref{tb:brlist} 
labels rows of this type as `Flux'.

The number of parameters depends on the 
number of resonances to be fit and
the number of open channels included in the fit.  For example, the 
S$_{11}$ partial wave
has three resonances and we fit values for 
each bare pole and its couplings to as many as 8 channels.  
There are also two subthreshold and one high energy ``states''
used to simulate background.   
The masses of these states are constrained to be
far away from any of the actual resonances.
Three parameters, a pole and coupling strengths
to $\pi N$ and $\eta N$ are fit for each subthreshold background pole.
The high energy state is allowed to couple to all open channels.  
Thus, there are 38 parameters fit in this partial wave.  
For the D$_{15}$ partial wave, only one resonance (with 
coupling to four channels) 
and three background poles were fit, a total of 15 real parameters.  

In the elastic channel, the data quality is 
reasonable at values of W from threshold to about 2.0 GeV.  
At values of W larger than roughly 1.8 GeV, no inelastic data
is available for the $\pi \pi N$ final state.  
The inelastic data has significant 
fluctuations.  If the data are to be represented by a smooth function 
(assumed in all analyses), the 
error bars are underrepresented.  
In fact, the Manley {\it et al.} paper~\cite{Manley84}
states that only diagonal errors were included in their output.  
Correlations should be significant in the analysis 
and can only add to the uncertainties.  Although the elastic
data were able to be fit well, the inelastic data were not.  
Since the elastic data is of much higher quality than the inelastic
data, the inelastic error bars were weighted 
by a factor of 2 lower than the elastic
error bars in order to ensure a reasonable fit to the elastic data.  
(The original Cutkosky, et al. paper~\cite{Cut79,Cut90} used a factor 
of 3 to weight the inelastic data.)  
For elastic data, values of $\chi^2/data point$ 
were 1.7 and 1.6 for the S$_{11}$ and D$_{15}$ waves, respectively.  
For the inelastic amplitudes, $\chi^2/data point$ 
values were 9.2 and 22.2.  The $\chi^2$ values are given for the
partial wave amplitude values without the extra weighting factor.  
(We don't quote $\chi^2$ per degree of freedom because 
the parameters are shared between the elastic and
inelastic fits.)  
Although we get qualitatively better fits to the elastic 
data than Manley and Saleski\cite{ManSal} in most cases, 
similar quality fits are found for the inelastic 
amplitudes.  (No values of $\chi^2$ are given by KSU.)  
However, the shapes of our inelastic $T$ matrices are qualitatively 
different in many cases.   Specific partial waves will be discussed in
section~\ref{se:detailedresults}.

	For a complicated multi-parameter fit, errors
are difficult to determine because correlations can be significant.  
The partial wave data we use as input quotes only diagonal errors.
We include error estimates in all extracted quantities due to
propagation of errors quoted in the partial wave data.  In addition, 
we add contributions
determined from additional fits where the background parameterization
is varied.  

	To allow a full error analysis, separate fits were made to data over a 
limited energy range to isolate single resonances.  A $2 \times 2$ $K$ matrix was
used to model 2 channels at a time and a number of fits were made
for each resonance to determine errors on each of the extracted
quantities.  In one fit, the errors on the resonance mass and width
and the error on the largest branching ratio were determined.
With $\pi N$ as one channel
and the channel with the largest remaining branching fraction 
(often $\pi \Delta$) for the resonance under consideration as the second
channel, the resonance mass, width, and branching fraction together 
with a simple
parameterization of the nonresonant amplitude were fit to the data.
Errors for the branching fraction to the 
channels with less coupling strength were determined in fits where the 
mass and width of 
the resonance were fixed at the final fit value of this work and only 
the branching fraction and background were fit.  For the $K$ matrix in
these cases, the channel for which the branching fraction was being
determined was one $K$-matrix channel and the $\pi N$ channel or 
the channel with the
largest branching fraction was chosen as the second K-matrix 
channel.  In all cases, two fits
were done with different simple assumptions for the background dependence,
either flat vs. linear or linear vs. quadratic.  The first type was most 
often used.  Two components to the error for each parameter under study 
were determined from each 
pair of fits, a {\it relative error} for each fit parameter from the fit 
with the largest $\chi^2$ 
and the absolute difference between the two determinations of the fit 
parameter.  The first component is representative of the 
statistical error in the data points; the second error is due to systematic
effects between the different background parameterizations.  
When two reasonable
fits were obtained, the components were added in quadrature.  For prominent
states, the fits were easily made.  However, most states required many 
trials to find fits of the appropriate quality.  Background shapes, 
channel choices, W ranges, and weighting of the inelastic channel 
in the $\chi^2$ determination were all varied until two good fits 
were obtained.  All error bars quoted in
the tables discussed below were determined from these fits.

\subsection{General Results}

A list of resonances found in this analysis is given in Tables 
\ref{tb:reslist} and \ref{tb:brlist} and compared to
the results of KSU~\cite{ManSal}, those of CMB~\cite{Cut79} and the latest 
recommended values given by PDG~\cite{PDG}.  For various states, many
other results exist.  We make direct comparisons only with the
previous results that use multi-channel models and provide error
bars for their determinations.
We show masses and widths in Table \ref{tb:reslist} 
and branching fractions to 
various inelastic channels in Table \ref{tb:brlist}.  
These are the results of the analysis discussed in sect.~\ref{se:respar}.
Poles corresponding to most of the states found by KSU are found in this 
analysis, although the properties can be different; states for which there
is no evidence in the present analysis 
are P$_{13}$(1900),  F$_{15}$(2000), and D$_{33}$(1940), all of which
are 1* states according to PDG~\cite{PDG}.
Although we obtain fit results for many weak states, 
a detailed study of the 
validity of the fit for weak states was not attempted.  
Discussion in this section 
will be limited to PDG 3* and 4* resonances.
In Table IX, we show the results for pole positions 
of this analysis.  The pole positions have less 
model dependence than the parameters in Tables~\ref{tb:reslist}
and \ref{tb:brlist}.

	The analysis presented here is not identical to any previous
analysis.  The data base used is similar to KSU, but the formalism has 
significant differences.  The formalism is identical to CMB, but the 
data base used is quite different.  CMB used the best representation of
data at the time- their own elastic amplitude 
analysis\cite{Cutpin} and the
inelastic quasi-two-body amplitudes of the SLAC-Berkeley\cite{SLACpipi} 
and Imperial College\cite{Imppipi}.  Batinic {\it et al.}\cite{Bat} 
use a truncated version
of the formalism used here and still another data set.  

Strong isolated resonances that have a strong elastic coupling are  
fit well with all models and results for the resonance parameters,
such as the D$_{15}$(1675) and F$_{15}$(1680) masses and widths, tend to
have close agreement between previous results and the new results.  PDG
gives a range of 15 MeV for both masses and a
range of 40 and 20 MeV for the D$_{15}$ and F$_{15}$ full widths and our 
values are within these intervals.  For the elastic branching fraction, PDG
suggests a range of 10\%; our results are inside the range for F$_{15}$
and just outside it for D$_{15}$.

The benefit of the multichannel analysis 
is readily apparent for states
with a very small elastic branching fraction.  For example, D$_{13}$(1700) 
and P$_{11}$(1710) (both PDG 3* resonances) are
not seen in the VPI elastic analysis~\cite{VPIpin} 
because there are no strong 
signs of the resonance in the elastic $T$ matrix.  However,
there is a strong resonance signal in the $\pi N \rightarrow \pi \Delta$
T matrix in each case.  

For the cases where KSU differs significantly 
from the consensus of previous results, 
the S$_{31}$(1620) mass and the S$_{11}$(1650) elasticity, the new results
tend to agree with the older values.  To test for the dependence on
the elastic data used in the fit,
we re-did fits with the same elastic data used by KSU and found
results qualitatively similar to our final analysis.

	An unusual feature of the CMB analysis
is the large value for the S$_{11}$(1535) width, about 40\% larger than
any other analysis.  This analysis obtains a width in closer agreement with
KSU and KH than CMB.  We have been unable to reproduce the large
width.  We note that CMB based their fit on a subsidiary analysis
of Bhandari and Chao~\cite{CMBetan} with a compatible model.  Although
the $\pi N \rightarrow \eta N$ data has changed little since then, the
elastic data we use is a global fit to all data rather than the single 
data set they used.  The VPI SM95 single energy solution is systematically 
larger and somewhat flatter than 
the data used in that 1977 work.  The most confusing aspect is that
Bhandari and Chao quote a full width of 139$\pm$33 MeV, much more 
compatible with the present results than with the full CMB results.

There has been significant interest and controversy in the properties 
of states in the P$_{11}$ partial wave.  In large part this is because the 
data in this partial wave has always been poor.  In the present fit,
$\chi^2/data point$ for this partial wave is 3.8 for the elastic 
channel and ranges from 8.5 to 17.6 for the inelastic channels.  
With the low quality of the existing inelastic data, no effort was 
made to determine the correct number of P$_{11}$ states.  
We find three P$_{11}$ states
with properties somewhat different than those previously obtained,
although all of our values are within the suggested ranges of PDG.  
In our results, P$_{11}$(1440) has a mass that just overlaps
the PDG window and one of the largest widths obtained.  In our fit, 
this state has a significant contribution from nonresonant 
interactions; that
together with the low quality of the present data 
produces a large estimated 
error for the values given.  KSU results for this state show a somewhat 
smaller width and a much smaller error bar.  We can only comment that 
the KSU model handles background quite differently than the present 
model and they used older $\pi N$ elastic data.  For the P$_{11}$(1710),
we find mass and width values well within the large PDG ranges; however,
KSU has a very large width with a large estimated error.  Since 
this state sits on the tail of the P$_{11}$(1440) and doesn't have 
a strong signal in any channel, it is clear
that the properties of the 2 lowest P$_{11}$ states are closely
coupled and multichannel analyses are very appropriate.  A later analysis of
Cutkosky and Wang~\cite{Cut90} of this partial wave using the CMB model 
compared results obtained with the VPI (SM89) and 
CMB80 partial wave analyses.  They find qualitatitively similar results
(large width for the Roper and small width for the 1700 MeV state) to
those obtained here.  Evidence for 
the highest energy P$_{11}$ state in the 
present data is very poor because only elastic data exists 
in the appropriate energy range.

Other unusual cases found in this new analysis include the
P$_{31}$(1910) and S$_{31}$(1900) states.  
The P$_{31}$(1910) is found at significantly higher mass and has
significantly larger width than previous determinations. 
Since this state is at high mass, the inelastic data is very important
in determining its properties.  However, there is almost no existing
inelastic data in this partial wave other than a few points in
the $\pi N \rightarrow \pi N^*(1440)$ reaction.  PDG has given this
state a 4* rating, but with the present data this rating should be 
downgraded.  Although VPI doesn't find 
the S$_{31}$(1900) state, it is very prominent in KSU and PDG.  
We find the mass at 1802 MeV
and a width of 48 MeV (with a large estimated error) while KSU finds
1950 MeV and 263 MeV.  This significant difference is largely due to
the elastic data sets used.  There
is a strong bump at about 1900 MeV in the elastic $T$ matrices used by KSU
which has vanished in the VPI partial wave amplitudes. 

For the branching fractions presented in Table \ref{tb:brlist}, the only
recent result is KSU, which is presumably weighted heavily in the PDG 
listings (also shown in the table).  
We divide the $\rho N$ and $\pi \Delta$
channels into the appropriate spin channels since the component 
angular momenta can sometimes
have more than one value.  For D$_{13}$, the $\rho N$ spin can be 
1/2 with orbital angular momentum 2 or 3/2 with orbital angular
momentum 0 or 2.  Since only the spin 3/2 orbital angular momentum 0 case
was found to be important in the $\pi N \rightarrow \pi \pi N$
isobar analysis~\cite{Manley84}, this is the only
$\rho N$ channel we include for this partial wave.
The $\pi \Delta$ channel can couple with orbital angular momentum of 0
or 2 in this partial wave and both possibilities are included in the fits.  

As with KSU, uncertain fits due to underestimated error bars 
and/or missing data make interpretation difficult in some cases.  
We quote 
values for this analysis and give estimated errors for each quantity.  

Since we use the same inelastic $T$ matrices as KSU for input to the fit, 
the results should
be qualitatively similar.  Based on our study in
section~\ref{se:modeldep}, we feel there is roughly 20\% difference
between the KSU and Pitt-ANL results due to model dependence
in the most sensitive quantities.  
The poor fits make this true less often than might be expected.  
For F$_{15}$(1680), the elastic branching fraction is $\sim$ 70\% in both
analyses and the largest inelastic channels are $\pi \Delta$ P-wave and
$(\pi \pi)_s N$ in both; agreement is within errors for the largest values.
For D$_{33}$(1700), the elastic branching fraction is small and the 
inelastic strength
in concentrated in the $\pi \Delta$ s-wave, so we are in close agreement
with KSU.  However, S$_{31}$(1620) is a strong state where agreement
is not good.  There are 2 strong inelastic channels, $\rho_1 N$ and $\pi 
\Delta$ D-wave.  Although there is agreement in the strength of $\rho_1 N$,
the elastic strength is much smaller for KSU and the 
$\pi \Delta$ branching fraction is
of course larger.  The full widths seen are only 15\% different and the
estimated errors overlap.  The $\chi^2$ per data point for
the S$_{31}$ partial wave in the elastic channel is 2.1 and ranges
from 4.3 to 17.2 for the inelastic channels.

   The calculation of Capstick and Roberts uses a relativized quark model
to calculate resonance masses and a $^3P_0$ model for creation of 
$q\bar{q}$ pairs~\cite{CapstickRoberts}.  
They fit the two quark level coupling parameters to 
the $\pi N$ decay amplitudes of the non-strange resonances rated by
PDG as 2* or better and then predict the remaining decay amplitudes
for various meson+baryon final states (including many final states
for which there are no data).  Although their primary purpose was to
look for signatures of ``missing'' quark model resonances, we note that
the qualitative agreement with our analysis is satisfactory.  There are
notable successes such as D$_{13}$(1520) where quantitative agreement
comes with all significant decay channels and notable failures such as 
S$_{11}$(1535) where both $\pi N$ and $\eta N$ decay widths (and the 
full width) are overestimated by a factor of 4.

\subsection{Detailed Discussion - D$_{15}$, D$_{13}$, and S$_{11}$ 
Partial Waves}
\label{se:detailedresults}

	Presented in this section are  
detailed results for three representative partial waves.
It includes figures of all the channel $T$ matrices 
and a discussion of the results.  These partial waves have 
reasonable quality data and
contain both prominent and less prominent states.  
Numerical data can be found in
Tables \ref{tb:s11mod1}, \ref{tb:s11mod2}, \ref{tb:brlist}, and 
\ref{tb:reslist}.
The $T$ matrices are found in figures
\ref{fig:D15} for the D$_{15}$ partial wave,
\ref{fig:D13} for D$_{13}$, and \ref{fig:S11} for S$_{11}$.  In
each we show the relevant phase shift 
amplitudes~\cite{VPIpin,Manley84} along with the fit of Manley and 
Saleski~\cite{ManSal} (dashed lines) and our final fit (solid line).
Only data up to W=2.15 GeV were used in the fits because the data
at higher values of $W$ are of diminished quality.  Since all the data 
in each partial wave were simultaneously fit with the requirements
of unitary, not all channel data shown in the figures is fit 
equally well.
In general, the fits to the elastic data are good ($\chi^2 / datapoint 
\sim 2$) and the fits to
the inelastic data are poor ($\chi^2 / datapoint > 10$) for both
the KSU fits and the present results, as discussed above.

  The D$_{15}$ partial wave contains a strong (PDG\cite{PDG} 4*),
isolated resonance for which the results of the one resonance 
example (see section~\ref{se:singres})
apply fairly well.  In this case, both the elastic and 
$\pi N \rightarrow \pi \Delta (D wave)$ inelastic channels are
prominent in the D$_{15}$ partial wave.  Since neither threshold is nearby 
and there is no interfering
resonance, the $T$ matrix in each of these channels has the characteristic 
shape of an isolated resonance with a peak in the elastic and inelastic
cross section near the resonance mass.  The non-resonant amplitude is a 
smooth function.  The KSU and present 
fits are similar in the ability to match features in 
the data in the two strong channels near the cross section peak
at $W\simeq$ 1680 MeV.   The two analyses differ in their description
of the high energy data ($W\geq$ 1.8 GeV), but the data are very
sparse there.  Neither model fits the smaller $\rho$ production channels
well.  The $\rho_3 N$ channel shows that the 
model expectations (based on the 
requirements of unitarity and the data in the other channels) don't 
match the data in this channel and a poor fit results.

	Essentially all analyses for the D$_{15}$(1680) 
resonance give similar values for the mass, full width, 
and elastic fraction 
(see Table~\ref{tb:reslist}).  
Even though VPI~\cite{VPIpin} accounts for 
the inelasticity by using a dummy channel, the fit parameters for
this state are similar to those obtained in the present fit.  
PDG~\cite{PDG}
gives quite small error bars for the mass and width of this state, 
reflecting the unanimity of the fitting results for this state.  It 
is encouraging
that the complicated features of the present model are not important 
for the simple case presented in this partial wave.  

	The D$_{13}$ partial wave contains a 4* state at 1520 MeV (PDG
standard value, the actual mass is slightly different), a 3* state 
at 1700 MeV, and a 2* state at 2080 MeV.  Since the lowest two resonances
have significant overlap in energy, the features of the CMB model are
important for this partial wave.
The lowest state is highly elastic, but also shows up prominently in
$(\rho_3 N)_S$ and $(\pi \Delta)_S$ final states.  Note that the peak in
the imaginary $T$ matrix is inverted.  As already mentioned, the second
state is barely seen in the elastic channel, thus is not found in the 
VPI~\cite{VPIpin} analysis.  Both KSU and the present analysis
find most of the decay strength in the $\pi \pi N$ final states with 
less than 10\% of the decay strength to the elastic channel.  KSU differs
with this model in the distribution of the inelastic strength.  While
KSU finds most of the strength in $\rho N$, this model finds it in
$\pi \Delta$.  The KSU model fits available inelastic data better.  
Feuster and Mosel~\cite{Feuster} find a systematically lower width than
other models for the lowest state, but their width for the 2nd state is 
unusually large.  Evidence for the third state is very
weak in this analysis; there is only elastic data at $W\geq$ 1.9 GeV and
there is very little evidence of structure.  It is best fit with a very
large width.

	The 3 resonances have varying strength in the different
channels.  As a result, the fitted curve also changes significantly
from channel to channel.  In the $\pi \Delta$ channels, the
shapes are complicated because the 1700 MeV state is important.  The imaginary
part of the $T$ matrix has a dip at the 1520 MeV state and a peak at the 1700
state.  Nevertheless, this model and the KSU model fit the features well
and the fitted parameters for the higher energy state agree within 
stated errors.

  The S$_{11}$ partial wave was already discussed in
section \ref{se:s11} with regard to the significant model
dependence seen.  As a result of this model dependence, the 
S$_{11}$(1535) full width has 
been quoted as 66-250 MeV based on different analyses
of similar data (almost all fits are based prominently on the elastic 
$\pi N$ channel).  The relatively small width of the 
S$_{11}$(1535) in this analysis
(112$\pm$30 MeV) is determined in large part by a significant
overlap of the S$_{11}$(1535) with the S$_{11}$(1650).
This overlap causes a large interference effect.  Although a similar 
effect was seen in one description of the 
$\gamma p \rightarrow \eta p$ data that uses a formalism similar
to the Pitt-ANL model~\cite{Sauermann}, many models based on a $T$-matrix
formalism (e.g. \cite{BenMuk95,TiatBenh,Suryapin}) have a formalism 
where it is difficult to account for more than one resonance in a partial 
wave.  

The multichannel $K$-matrix analysis 
of Feuster and Mosel~\cite{Feuster} gets values of the full width in the 
range of 151-198 MeV for different form factors.  However, their model
doesn't handle the $\eta N$ cusp as well because it lacks the analytic
phase space, doesn't model the $\pi \pi N$ inelasticity well because a
zero width particle is used, and doesn't 
include off-shell intermediate state scattering effects because of the
$K$-matrix approximation.

  Although 8 channels are included in the fit, 
many of them turn out to have
small coupling to S$_{11}$(1535) (in agreement with PDG~\cite{PDG}).  
The two major channel couplings for S$_{11}$(1535) are 
$\pi N$ and $\eta N$.  Therefore, another determining 
factor in the full width is the total cross section
for the $\pi^- p \rightarrow \eta n$ in Fig.~\ref{fig:pietaxs} and the
corresponding $T$ matrix in Fig.~\ref{fig:S11}b.  
The cross section of Fig.~\ref{fig:pietaxs} is made up of 
$\pi N \rightarrow \eta N$ $T$-matrix amplitudes for S$_{11}$ through
G$_{17}$, however its main component is S$_{11}$.
Although the 
shape of the cusp (see Fig.~\ref{fig:S11cusp}) does carry information 
about the $\eta N$ channel, the overall data quality isn't satisfactory.  We
await higher quality data for a more conclusive determination of this
amplitude.

	The branching fraction of the S$_{11}(1535)$ to
the $\pi N$ channel is at the low end of the PDG range.
A surprising feature of the
S$_{11}$(1535) has been its unusually large decay width
to $\eta N$.  We find strong evidence for a value at the
high end of the PDG range.
We believe our result because our model fully accounts for
the threshold enhancement (cusp effect) due to the 
$\eta N$ channel opening and the interference of the 2 overlapping
resonances.  Since no other model has such a complete formulation, 
they can have model-dependent systematic errors as a result.  

  The properties of S$_{11}$(1650) are largely determined 
through the prominent bump in the elastic channel.  The
elastic $T$ matrix has a peak in the imaginary part at about
the right energy and the real part is decreasing at that energy;
a strong resonance signal.  The strongest inelastic signal comes
in the $\rho_3 N$ channel.  The decay width to $\eta N$ is small 
in this analysis, in agreement with previous analyses.  The striking
difference between branching fractions to $\eta N$ has been another
unusual feature of these states.  The existence of the third S$_{11}$ 
resonance is very weakly supported by the existing data.  In our
analysis, this state is most often determined by the need to
fit non-statistical features in the partial wave amplitudes.
Further data will be required to sort out the question of whether 
or not it exists.  

	The role of the inelastic $\eta N$ channel is very important in the
determination of resonance parameters for the S$_{11}$(1535) state,
as shown in some detail in section~\ref{se:s11}.  Unfortunately, the 
existing data for $\pi N \rightarrow \eta N$ is sparse and of uncertain
quality~\cite{Clajus} (see Fig.~\ref{fig:pietaxs}).  Recently published 
data~\cite{Krusche} for  
$\gamma p  \rightarrow \eta N$ has been interpreted as evidence for
a very large width ($\simeq$ 200 MeV).  We do not include the Krusche
{\it et al.} data in this analysis.  However, very
simple resonance models~\cite{Krusche,Nimai} were used to determine 
the large width.  Since the large body of $\pi N$ data was not fitted with
their model, possible inconsistencies exist in their parameterization.  
Although the simple model has validity for the $\eta N$ coupling to 
S$_{11}$(1535), we feel there will still be significant effects from 
interference with S$_{11}$(1650) and coupled channel effects (e.g. 
$\gamma N \rightarrow S_{11}(1650) 
\rightarrow \pi N \rightarrow S_{11}(1535)
\rightarrow \eta N$).

\subsection{Observables}
$T$-matrix amplitudes were fit using the Pitt-ANL model.  
We do not fit directly to data; rather, we fit the partial wave 
amplitudes.  It is then interesting to compare the 
actual experimental data to our calculated observables
to verify that nothing has been lost by not fitting to the 
actual data points.

The $f$ and $g$ spin non-flip and spin flip amplitudes are defined
in terms of the Pitt-ANL $T$-matrix elements for $\pi N$ elastic scattering
by~\cite{Goldberger-Watson}:
\begin{equation}
\begin{array}{rcl}
f(\theta ) & = &
\frac{1}{q}\sum_{\ell =0}^\infty \left[ (\ell + 1)T_{\ell +} + 
                                          \ell T_{\ell -}       \right]
                                 P_\ell (\cos{\theta}) \\
g(\theta ) & = &
\frac{1}{q}\sum_{\ell =1}^\infty \left( T_{\ell +} - T_{\ell -} \right)
                                 P_\ell'(\cos{\theta})
\end{array}
\label{eq:fgofT}
\end{equation}
where $P_\ell$ and $P_\ell'$ are the Legendre polynomials and their first
derivative with respect to $\cos{\theta}$, respectively.  The $T_{\ell\pm}$ are
the amplitudes for a particular $\pi N$ scattering charge channel, where
the $\ell\pm$ refers to the total spin $J=\ell\pm \frac{1}{2}$, and $q$ is
the center of mass momentum of the final state pion.  The $T_{\ell\pm}$ are
defined by Eq. (\ref{eq:tcutkosky}) for
the $\pi N \rightarrow \pi N$ channel.

The differential cross section $(\frac{d\sigma}{d\Omega})$ and polarization P 
are:
\begin{equation}
\begin{array}{rcl}
\frac{d\sigma}{d\Omega}   & = & |f(\theta )|^2 + |\sin{\theta}g(\theta )|^2\\ 
\frac{d\sigma}{d\Omega} P & = & 2 \Im m \left[ f(\theta ) g^*(\theta ) \right]
                               \sin{\theta}
\end{array}
\label{eq:pipiobs}
\end{equation}

Figure~\ref{fig:cxs_totxs} shows a plot of the total cross section 
vs. the Pitt-ANL calculation for the
process $\pi^- p \rightarrow \pi^0 n$.  Figure~\ref{fig:pipp_dsg}
shows plots of the differential cross section, $\frac{d\sigma}{d\Omega}$,
at a number of different energies for the process
$\pi^+ p \rightarrow \pi^+ p$.  
Figure~\ref{fig:pimp_pol}
shows plots of the nucleon recoil polarization, P,
for a number of different energies for the process
$\pi^- p \rightarrow \pi^- p$. 
In each case, the calculation using 
final fit parameters is in good agreement with the data. 
For the P data, the error bars are sometimes large.

For the plot in 
figures~\ref{fig:cxs_totxs}, \ref{fig:pipp_dsg}, and \ref{fig:pimp_pol} the
data is taken from VPI's SAID database.  It is a large collection
of scattering data for a number of different reaction channels,
including $\pi N$ elastic and charge exchange reactions.  

Based on our studies, we can conclude that our resonance parameters
are very close to what would have been obtained if the fit had been
made to the original data.


\section{Conclusions}
We have presented a new study of excitation of baryon resonances
through pion-nucleon interactions.  
An expanded version of the Carnegie-Mellon
Berkeley (CMB) was used in this work.  We added the $\eta N$ channel to
the calculation and used the latest data sets.  The model satisfies
many of the desirable theoretical constraints- two 
and three body unitarity,
time reversal invariance, and analyticity.  Because the model is based
on a separable interaction, it is linked to a Hamiltonian approach,
unlike many of the models previously used~\cite{VPIpin,ManSal}.  The 
model has a key feature of allowing large numbers of asymptotic channels and
more than one resonance per partial wave.  Since analyticity is satisfied
and all inelastic thresholds are included,
structure in the physical observables due to 
nonresonant effects are carefully
treated.  Resonances can then be 
found as poles in the complex $s$ plane.
As in the original CMB model,
non-resonant processes are included as very wide resonances at very
high or very low energies which are then subject to the same threshold
effects as the resonances.  This empirical approach is well established
in potential scattering theory, but could be 
replaced later by theoretical calculations based on meson exchange 
models, e.g.~\cite{SatoLee}. 

Rather than deal with the complexities of a 
direct fit to the cross section and polarization data, we use
the partial wave amplitudes of the VPI and KSU 
groups~\cite{VPIpin,Manley84} and our own fit to the
$\pi N \rightarrow \eta N$ data.
A re-evaluation of the $\pi N \rightarrow \pi \pi N$ 
data is in progress by our group.  

The features of this model are expected to be very 
important in the S$_{11}$ partial wave
where there are two overlapping resonances 
at 1535 and 1650 MeV and a strong threshold opening.  The lower
state pole and the $\eta$N channel opening close together 
and the quantum mechanical interference of the two states is found
to be significant.  We show how the interpretation of these states has
very strong model dependence and how relaxing the features of the
present model can skew the fitting results for the width and branching 
fractions to $\pi N$ and $\eta N$.  We get a new result for the
branching fraction to $\eta N$ at the high end of what has been
previously published for S$_{11}$(1535), making its characteristics 
even more unusual.

The general fitting results are presented in Tables~\ref{tb:reslist} and
\ref{tb:brlist}.  
We searched for the same states seen in the previous analysis
of Manley and Saleski~\cite{ManSal}.  This simplification was made because
the data quality wasn't high enough for a valid search for weakly excited 
states.
Although values of $\chi^2/data point$ are roughly 2 for elastic fits, they
are 10 or more for fits to the $\pi N \rightarrow \pi \pi N$ $T$ matrix data.  
This is perhaps due
to the lack of a treatment of correlations in the error analysis
of the pion production data.  Modern experimental techniques could greatly 
improve the inelastic data set.  We strongly encourage new measurements of 
the inelastic channels.  A significant effort was made to determine error
bars reflecting both the estimated errors in the $T$ matrices used in the
fits and the differing choices of background energy dependence.

We find strong and isolated states 
(e.g. P$_{33}$(1232) and D$_{15}$(1675))
with very similar parameters as previous 
analyses.~\cite{VPIpin,ManSal,KH80,Cut79}.
States with strong model dependence such as 
S$_{11}$(1535) or with significant
changes in the data set such as S$_{31}$(1900) and D$_{13}$(1700) get
quite different results.  We find a full width of S$_{11}$(1535) at the low
end of the range of previous values, especially 
with respect to the interpretations
of recent $\gamma p \rightarrow \eta p$ data~\cite{Krusche}.  The results 
presented here do not include the photoproduction data; rather, the 
present results are
dependent on a fairly weak data set for $\pi^- p \rightarrow \eta n$.  
Nevertheless, the large widths obtained in
analyses highly dependent on the Mainz eta photoproduction data are
only loosely coupled to the large data set used in this study.  

We are in the process of further extending the 
model to include photoproduction and electroproduction reactions.  
The Born terms are included for production
of pions and etas and the resonance spectrum is being refit.  
Results of the present paper will then be subject to change.

\section{Acknowledgments}
We are grateful to R. Arndt, R. Workman, and M. Manley for
sharing their partial wave data and their analysis results
with us.  We have also benefited greatly from the programming and
database work of D. Ciarletta, J. DeMartino, J. Greenwald, D. Kokales, 
M. Mihalcin, and K. Bordonaro at the University of Pittsburgh.
Special thanks go to C. Tanase for investigating the analytic
structure of this model in detail.

\begin{table}
\caption{Model parameters for a {\it test} calculation in the 
S$_{11}$ partial wave.  
These are the bare mass, the physical mass and width, 
and the dressed pole location for the 
two resonances included in the calculation.  No
background contributions are included.}
\label{tb:s112by2m}
\begin{tabular}{ccccc}
state          & bare    & physical    & full      & physical \\
               &  mass     & mass   & width    & pole   \\
               & (MeV)     & (MeV)   & (MeV)   & (GeV) \\
\hline
S$_{11}(1535)$ & 1500     &   1518  &  83      & 1.507-0.032i \\
S$_{11}(1650)$ & 1650     &   1680  & 200      & 1.676-0.101i  

\end{tabular}

\end{table}
\begin{table}
\caption{Model parameters for {\it test} calculation in the 
S$_{11}$ partial wave.  
These are
the coupling parameters and branching fractions (b.f.) between the two 
asymptotic channels and the
two resonances included in the calculation.  No
background contributions are included.}
\label{tb:s112by2br}
\begin{tabular}{ccccc}
state   & $\gamma_{\pi N}$ & $\gamma_{\eta N}$ & $\pi N$  & $\eta N$  \\
        &  (GeV)         &    (GeV)        &  b.f.     &   b.f.   \\
        &                &                 & (\%)      &    (\%)   \\
\hline
S$_{11}(1535)$ &     0.5      &     0.85       &   42     &      58  \\
S$_{11}(1650)$ &     0.9      &    -0.04       &   96     &      4      

\end{tabular}
\end{table}

\begin{table}
\caption{Channels included in this analysis.  Since isospin symmetry is
assumed to simplify the analysis, all charge states of a particle
have the same mass.  We choose channels very similar 
to those used by KSU with identical nomenclature.  All 
channels used in fits have fixed spin, orbital, and total angular 
momentum.  In the text, the $\rho N$ channel is denoted by 
$(\rho_s N)_{\ell}$ with $s$ equal to
twice the total spin of the $\rho N$ system and $\ell$ giving the orbital
angular momentum in spectroscopic notation.  The orbital angular momentum 
of the $\pi \Delta$ is also ambiguous, 
so it is also given as a subscript.} 
\label{tb:chans}
\begin{tabular}{ccccc}
channel & baryon & baryon  & meson & meson \\ 
        & mass   & width   & mass  & width \\
\hline
$\pi N$   &   939  &         & 139   &   \\
$\eta N$  &   939  &         & 549   &   \\
$\rho N$  &   939  &         & 770   & 153  \\
$\sigma N$ &   939  &         & 800   & 800  \\
$\omega N$ &   939  &         & 782   &   \\
$\pi \Delta$ & 1232 &  115   & 139   &   \\
$\pi N*$  &  1440  &   200   & 139   &   \\
$K \Lambda$ & 1116 &         & 498  &   
\end{tabular}
\end{table}

\begin{table}
\caption{Model dependence for the S$_{11}(1535)$.  See text for details.}
\begin{tabular}{@{\extracolsep{\fill}}|cccc|ccccc|}
Unitarity & Disp. & Res. & Channels & 
Mass & Width & $\pi$N & $\eta$N  & $\pi\pi$N  \\
     &  Rel. & Type & in fit & MeV  & MeV   & \% & \% & \% \\
\hline
 K-Matrix & NO & NRBW & $\pi$N & 1518 & 87 & 43 & 6 & 51\\
 K-Matrix & NO & NRBW & $\pi$N , $\eta$N & 1532 & 108 & 45 & 39 & 16\\
 K-Matrix & NO & NRBW & ALL & 1535 & 126 & 42 & 44 & 14\\
 K-Matrix & NO & RBW & $\pi$N & 1514 & 84 & 35 & 0 & 65\\
 K-Matrix & NO & RBW & $\pi$N , $\eta$N & 1533 & 110 & 44 & 40 & 16\\
 K-Matrix & NO & RBW & ALL & 1534 & 125 & 42 & 43 & 15\\
 Dyson eq. & YES & RBW & $\pi$N & 1531 & 72 & 16 & 62 & 22 \\
 Dyson eq. & YES & RBW & $\pi$N , $\eta$N & 1526 & 114 & 36 & 41 & 23 \\
 Dyson eq. & YES & RBW & ALL & 1542 & 112 & 35 & 51 & 14 \\
\end{tabular}
\label{tb:s11mod1}
\end{table}

\begin{table}
\caption{Model dependence for the S$_{11}$(1650).  See text for details.}
\begin{tabular}{|cccc|ccccc|}
Unitarity & Disp. & Res. & Channels & 
Mass & Width & $\pi$N & $\eta$N  & $\pi\pi$N  \\
     &  Rel. & Type & in fit & MeV  & MeV   & \% & \% & \% \\
\hline
 K-Matrix & NO & NRBW & $\pi$N & 1645 & 233 & 35 & 49 & 16 \\
 K-Matrix & NO & NRBW & $\pi$N , $\eta$N  &1689 & 225 & 67 & 31 & 2 \\
 K-Matrix & NO & NRBW & ALL& 1694 & 259 & 72 & 16 & 12\\
 K-Matrix & NO & RBW & $\pi$N & 1682 & 161 & 78 & 5 & 17\\
 K-Matrix & NO & RBW & $\pi$N , $\eta$N & 1692 & 233 & 75 & 15 & 10 \\
 K-Matrix & NO & RBW & ALL& 1690 & 229 & 65 & 25 & 10 \\
 Dyson eq. & YES & RBW & $\pi$N & 1692 & 138 & 65 & 35 & 0\\ 
 Dyson eq. & YES & RBW & $\pi$N , $\eta$N & 1676 & 104 & 54 & 45 & 1 \\
 Dyson eq. & YES & RBW & ALL & 1689 & 202 & 74 & 6 & 20 \\
\end{tabular}
\label{tb:s11mod2}
\end{table}

\begin{table}
\caption{S$_{11}$ Resonance parameters from fits to different elastic
data sets.  All results use the $\pi N \rightarrow \eta N$ and 
$\pi N \rightarrow \pi \pi N$ partial wave amplitudes as in the full
analysis.  Comparisons are made for results using the VPI and
a mixture of the CMB80 and KH80 elastic partial wave amplitudes
(as was done in Ref.~[4]).  The first two lines in
the table give results for S$_{11}$(1535) and the last two
lines give results for S$_{11}$(1650).  Although all channels were used
in the fit, only the total $\pi \pi N$ branching fraction is given. }
\begin{tabular}{cccccc}
 Elastic &
 Mass  & 
 Width & 
 $\pi$N b.f. & 
 $\eta$N b.f. & 
 $\pi\pi N$ b.f. \\

  Data Set& MeV  & MeV   & \% & \% & \%  \\
\hline
VPI &       1542  &  112  &  35  &  51  &  14 \\
CMB/KH80 &  1535  &  137  &  35  &  53  &  12 \\
VPI &       1689  &  202  &  74  &   6  &  20 \\
CMB/KH80 &  1691  &  222  &  58  &  15  &  27 \\
\end{tabular}
\label{tb:elas_dep}
\end{table}



\begin{table}
\caption{Results for masses, full widths, and elastic fractions for 
all resonances
found in this analysis.  All resonances found in the KSU analysis were
searched for, but not all were found.  No attempt was made to find new
resonances because then data quality is not good enough for a new search.  
See text for more details.}
\label{tb:reslist}
\begin{center}

\begin{tabular}{@{\extracolsep{\fill}}ccccc} 
Resonance  &  Mass  & Width   &  Elasticity & Reference \\
           &  (MeV) & (MeV)   &  \%       &           \\
\hline
S$_{11}(1535)$ & 1542(3) & 112(19) & 35(8) & Pitt-ANL\\
***** & 1534(7) & 151(27) & 51(5) & KSU\\
 & 1520-1555 & 100-250 & 35-55 & PDG\\
 & 1550(40) & 240(80) & 50(10) & CMB\\
\\

S$_{11}(1650)$ & 1689(12) & 202(40) & 74(2) & Pitt-ANL\\
***** & 1659(9) & 173(12) & 89(7) & KSU\\
 & 1640-1680 & 145-190 & 55-90 & PDG\\
 & 1650(30) & 150(40) & 65(10) & CMB\\
\\

S$_{11}(2090)$ & 1822(43) & 248(185) & 17(3) & Pitt-ANL\\
** & 1928(59) & 414(157) & 10(10) & KSU\\
 & $\approx$ 2090 &  &  & PDG\\
 & 2180(80) & 350(100) & 18(8) & CMB\\
\\

P$_{11}(1440)$ & 1479(80) & 490(120) & 72(5) & Pitt-ANL\\
***** & 1462(10) & 391(34) & 69(3) & KSU\\
 & 1430-1470 & 250-450 & 60-70 & PDG\\
 & 1440(30) & 340(70) & 68(4) & CMB\\
\\

P$_{11}(1710)$ & 1699(65) & 143(100) & 27(13) & Pitt-ANL\\
**** & 1717(28) & 478(226) & 9(4) & KSU\\
 & 1680-1740 & 50-250 & 10-20 & PDG\\
 & 1700(50) & 90(30) & 20(4) & CMB\\
\\
P$_{11}(2100)$ & 2084(93) & 1077(643) & 2(5) & Pitt-ANL\\
** & 1885(30) & 113(44) & 15(6) & KSU\\
 & $\approx$ 2100 &  &  & PDG\\
 & 2125(75) & 260(100) & 12(3) & CMB\\
\\
P$_{13}(1720)$ & 1716(112) & 121(39) & 5(5) & Pitt-ANL\\
***** & 1717(31) & 383(179) & 13(5) & KSU\\
 & 1650-1750 & 100-200 & 10-20 & PDG\\
 & 1700(50) & 125(70) & 10(4) & CMB\\
\\


D$_{13}(1520)$ & 1518(3) & 124(4) & 63(2) & Pitt-ANL\\
***** & 1524(4) & 124(8) & 59(3) & KSU\\
 & 1515-1530 & 110-135 & 50-60 & PDG\\
 & 1525(10) & 120(15) & 58(3) & CMB\\
\\



D$_{13}(1700)$ & 1736(33) & 175(133) & 4(2) & Pitt-ANL\\
**** & 1737(44) & 249(218) & 1(2) & KSU\\
 & 1650-1750 & 50-150 & 5-15 & PDG\\
 & 1675(25) & 90(40) & 11(5) & CMB\\
\\

D$_{13}(2080)$ & 2003(18) & 1070(858) & 13(3) & Pitt-ANL\\
*** & 1804(55) & 447(185) & 23(3) & KSU\\
 & $\approx$ 2080 &  &  & PDG\\
 & 2060(80) & 300(100) & 14(7) & CMB\\
\\

D$_{15}(1675)$ & 1685(4) & 131(10) & 35(1) & Pitt-ANL\\
***** & 1676(2) & 159(7) & 47(2) & KSU\\
 & 1670-1685 & 140-180 & 40-50 & PDG\\
 & 1675(10) & 160(20) & 38(5) & CMB\\
\\

F$_{15}(1680)$ & 1679(3) & 128(9) & 69(2) & Pitt-ANL\\
***** & 1684(4) & 139(8) & 70(3) & KSU\\
 & 1675-1690 & 120-140 & 60-70 & PDG\\
 & 1680(10) & 120(10) & 62(5) & CMB\\
\\


F$_{17}(1990)$ & 2311(16) & 205(72) & 22(11) & Pitt-ANL\\
*** & 2086(28) & 535(117) & 6(2) & KSU\\
 & $\approx$ 1990 &  &  & PDG\\
 & 1970(50) & 350(120) & 6(2) & CMB\\
\\

G$_{17}(2190)$ & 2168(18) & 453(101) & 20(4) & Pitt-ANL\\
***** & 2127(9) & 547(48) & 22(1) & KSU\\
 & 2100-2200 & 350-550 & 10-20 & PDG\\
 & 2200(70) & 500(150) & 12(6) & CMB\\
\\

S$_{31}(1620)$ & 1617(15) & 143(42) & 45(5) & Pitt-ANL\\
***** & 1672(7) & 154(37) & 9(2) & KSU\\
 & 1615-1675 & 120-180 & 20-30 & PDG\\
 & 1620(20) & 140(20) & 25(3) & CMB\\
\\

S$_{31}(1900)$ & 1802(87) & 48(45) & 33(10) & Pitt-ANL\\
**** & 1920(24) & 263(39) & 41(4) & KSU\\
 & 1850-1950 & 140-240 & 10-30 & PDG\\
 & 1890(50) & 170(50) & 10(3) & CMB\\
\\

P$_{31}(1750)$ & 1721(61) & 70(50) & 6(9) & Pitt-ANL\\
** & 1744(36) & 299(118) & 8(3) & KSU\\
 & $\approx$ 1750 &  &  & PDG\\
\\

P$_{31}(1910)$ & 1995(12) & 713(465) & 29(21) & Pitt-ANL\\
***** & 1882(10) & 239(25) & 23(8) & KSU\\
 & 1870-1920 & 190-270 & 15-30 & PDG\\
 & 1910(40) & 225(50) & 19(3) & CMB\\
\\

P$_{33}(1232)$ & 1234(5) & 112(18) & 100(1) & Pitt-ANL\\
***** & 1231(1) & 118(4) & 100(0) & KSU\\
 & 1230-1234 & 115-125 & 98-100 & PDG\\
 & 1232(3) & 120(5) & 100(0) & CMB\\
\\

P$_{33}(1600)$ & 1687(44) & 493(75) & 28(5) & Pitt-ANL\\
**** & 1706(10) & 430(73) & 12(2) & KSU\\
 & 1550-1700 & 250-450 & 10-25 & PDG\\
 & 1600(50) & 300(100) & 18(4) & CMB\\
\\
P$_{33}(1920)$ & 1889(100) & 123(53) & 5(4) & Pitt-ANL\\
**** & 2014(16) & 152(55) & 2(2) & KSU\\
 & 1900-1970 & 150-300 & 5-20 & PDG\\
 & 1920(80) & 300(100) & 20(5) & CMB\\
\\

D$_{33}(1700)$ & 1732(23) & 119(70) & 5(1) & Pitt-ANL\\
***** & 1762(44) & 599(248) & 14(6) & KSU\\
 & 1670-1770 & 200-400 & 10-20 & PDG\\
 & 1710(30) & 280(80) & 12(3) & CMB\\
\\


D$_{35}(1930)$ & 1932(100) & 316(237) & 9(8) & Pitt-ANL\\
**** & 1956(22) & 526(142) & 18(2) & KSU\\
 & 1920-1970 & 250-450 & 5-20 & PDG\\
 & 1940(30) & 320(60) & 14(4) & CMB\\
\\

D$_{35}(2350)$ & 2459(100) & 480(360) & 7(14) & Pitt-ANL\\
** & 2171(18) & 264(51) & 2(0) & KSU\\
 & $\approx$ 2350 &  &  & PDG\\
 & 2400(125) & 400(150) & 20(10) & CMB\\
\\


F$_{35}(1752)$ & 1724(61) & 138(68) & 0(1) & Pitt-ANL\\
** & 1752(32) & 251(93) & 2(1) & KSU\\
\\

F$_{35}(1905)$ & 1873(77)& 461(111)& 9(1) & Pitt-ANL\\
***** & 1881(18) & 327(51) & 12(3) & KSU\\
 & 1870-1920 & 280-440 & 5-15 & PDG\\
 & 1910(30) & 400(100) & 8(3) & CMB\\
\\

F$_{37}(1950)$ & 1936(4.5) & 245(12) & 44(1) & Pitt-ANL\\
***** & 1945(2) & 300(7) & 38(1) & KSU\\
 & 1940-1960 & 290-350 & 35-40 & PDG\\
 & 1950(15) & 340(50) & 39(4) & CMB\\
\\


\end{tabular}

\end{center}
\end{table}



\begin{table}
\caption{Results for decay branching ratios of all resonances found in this
analysis.  Fractions are expressed as a percentage of the full width found
in table~\ref{tb:reslist}. }
\label{tb:brlist}
\begin{center}

\begin{tabular}[t]{@{\extracolsep{\fill}}ccccc} 
Resonance  &  Channel  & Pitt-ANL   &  KSU  & PDG \\
\hline
S$_{11}(1535)$  & $\pi$N  &      35(4)  &      51(5)  & 35-55  \\
  & $\eta$N  &      51(5)  &      43(6)  & 30-55  \\
  & $\rho_1$N  &       2(1)  &       2(1)  & 0-4  \\
  & ($\rho_3$N)$_D$  &       0(1)  &       1(1)  &   \\
  & ($\pi\Delta$)$_D$  &       1(1)  &       0(0)  & 0-1  \\
  & ($\sigma$N)$_P$  &       2(1)  &       1(1)  & 0-3  \\
  & $\pi$N$^*$(1440)  &      10(9)  &       2(2)  & 0-7  \\
\\
S$_{11}(1650)$  & $\pi$N  &      74(2)  &      89(7)  & 55-90  \\
  & $\eta$N  &       6(1)  &       3(5)  & 3-10  \\
  & $\rho_1$N  &       1(1)  &       0(0)  & 4-14  \\
  & ($\rho_3$N)$_D$  &      13(3)  &       3(2)  &   \\
  & ($\pi\Delta$)$_D$  &       2(1)  &       2(1)  & 3-7  \\
  & ($\sigma$N)$_P$  &       1(1)  &       2(2)  & 0-4  \\
  & $\pi$N$^*$(1440)  &       3(1)  &       1(1)  & 0-5  \\
\\
S$_{11}(2090)$  & $\pi$N  &      17(7)  &      10(10)  &   \\
  & $\eta$N  &      41(4)  &       0(3)  &   \\
  & $\rho_1$N  &      36(1)  &      49(22)  &   \\
  & ($\rho_3$N)$_D$  &       1(1)  &       0(1)  &   \\
  & ($\pi\Delta$)$_D$  &       1(1)  &       6(14)  &   \\
  & ($\sigma$N)$_P$  &       2(1)  &       4(10)  &   \\
  & $\pi$N$^*$(1440)  &       2(1)  &      30(22)  &   \\
\\
P$_{11}(1440)$  & $\pi$N  &      72(2)  &      69(3)  & 60-70  \\
  & $\eta$N  &       0(1)  &   & 0-8  \\
  & $\rho_1$N  &       0(1)  &   &   \\
  & ($\pi\Delta$)$_P$  &      16(1)  &      22(3)  & 20-30  \\
  & ($\sigma$N)$_S$  &      12(1)  &       9(2)  & 5-10  \\
  & K$\Lambda$  &       0(1)  &   &   \\
\\
P$_{11}(1710)$  & $\pi$N  &      27(4)  &       9(4)  & 10-20  \\
  & $\eta$N  &       6(1)  &   &   \\
  & $\rho_1$N  &      17(1)  &       3(7)  & 5-25  \\
  & ($\pi\Delta$)$_P$  &      39(8)  &      49(10)  & 15-40  \\
  & ($\sigma$N)$_S$  &       1(1)  &       2(4)  & 10-40  \\
  & K$\Lambda$  &      10(10)  &      37(10)  & 5-25  \\
\\
P$_{11}(2100)$  & $\pi$N  &       2(1)  &      15(6)  &   \\
  & $\eta$N  &      61(61)  &   &   \\
  & $\rho_1$N  &       4(1)  &      27(79)  &   \\
  & ($\pi\Delta$)$_P$  &       2(1)  &      24(18)  &   \\
  & ($\sigma$N)$_S$  &      10(1)  &      32(71)  &   \\
  & K$\Lambda$  &      21(20)  &       2(6)  &   \\

\\
P$_{13}(1720)$  & $\pi$N  &       5(5)  &      13(5)  & 10-20  \\
  & $\eta$N  &       4(1)  &   &   \\
  & $\rho_1$N  &      91(1)  &      87(5)  & 70-85  \\
  & Flux  &       0(1)  &   &   \\
\\
\\
D$_{13}(1520)$  & $\pi$N  &      63(1)  &      59(3)  & 50-60  \\
  & $\eta$N  &       0(1)  &   &   \\
  & ($\rho_3$N)$_S$  &       9(1)  &      21(4)  & 15-25  \\
  & ($\pi\Delta$)$_D$  &      11(2)  &      15(4)  & 5-12  \\
  & ($\pi\Delta$)$_S$  &      15(2)  &       5(3)  & 10-14  \\
  & ($\sigma$N)$_P$  &       1(1)  &   & 0-8  \\
\\
D$_{13}(1700)$  & $\pi$N  &       4(1)  &       1(2)  & 5-15  \\
  & $\eta$N  &       0(1)  &   &   \\
  & ($\rho_3$N)$_S$  &       7(1)  &      13(17)  & 0-35  \\
  & ($\pi\Delta$)$_D$  &      79(56)  &      80(19)  &   \\
  & ($\pi\Delta$)$_S$  &      11(1)  &       5(10)  &   \\
  & ($\sigma$N)$_P$  &       0(1)  &       2(4)  &   \\
\\
D$_{13}(2080)$  & $\pi$N  &      13(2)  &      23(3)  &   \\
  & $\eta$N  &       0(2)  &   &   \\
  & ($\rho_3$N)$_S$  &       6(6)  &      26(14)  &   \\
  & ($\pi\Delta$)$_D$  &      17(10)  &      21(14)  &   \\
  & ($\pi\Delta$)$_S$  &      40(10)  &       3(7)  &   \\
  & ($\sigma$N)$_P$  &      24(24)  &      27(12)  &   \\
\\
D$_{15}(1675)$  & $\pi$N  &      35(2)  &      47(2)  & 40-50  \\
  & $\eta$N  &       0(1) &   &   \\
  & $\rho_1$N  &       0(1) &       0(0)  & 1-3  \\
  & ($\rho_3$N)$_D$  &       1(1)  &       0(0)  &   \\
  & ($\pi\Delta$)$_D$  &      63(2)  &      53(2)  & 50-60  \\
\\
F$_{15}(1680)$  & $\pi$N  &      69(1)  &      70(3)  & 60-70  \\
  & $\eta$N  &       0(1)  &   &   \\
  & ($\rho_3$N)$_F$  &       3(1)  &       2(1)  & 1-5  \\
  & ($\rho_3$N)$_P$  &       5(1)  &       5(3)  & 0-12  \\
  & ($\pi\Delta$)$_F$  &       1(1)  &       1(1)  & 0-2  \\
  & ($\pi\Delta$)$_P$  &      14(3)  &      10(3)  & 6-14  \\
  & ($\sigma$N)$_D$  &       9(1)  &      12(3)  & 5-20  \\
\\


F$_{17}(1990)$  & $\pi$N  &      22(3)  &       6(2)  &   \\
  & $\eta$N  &       0(1)  &      94(2)  &   \\
  & Flux  &      77(77)  &   &   \\
\\
G$_{17}(2190)$  & $\pi$N  &      20(1)  &      22(1)  & 10-20  \\
  & $\eta$N  &       0(1)  &   &   \\
  & ($\rho_3$N)$_D$  &      29(28)  &      29(6)  &   \\
  & ($\omega_3$N)$_D$  &      51(51)  &      49(7)  &   \\
\\
S$_{31}(1620)$  & $\pi$N  &      45(1)  &       9(2)  & 20-30  \\
  & $\rho_1$N  &      14(3)  &      25(6)  & 7-25  \\
  & ($\rho_3$N)$_D$  &       2(1)  &       4(3)  &   \\
  & ($\pi\Delta$)$_D$  &      39(2)  &      62(6)  & 30-60  \\
  & $\pi$N$^*$(1440)  &       0(1)  &   &   \\
\\
S$_{31}(1900)$  & $\pi$N  &      33(6)  &      41(4)  & 10-30  \\
  & $\rho_1$N  &      30(2)  &       5(7)  &   \\
  & ($\rho_3$N)$_D$  &       5(1)  &      33(10)  &   \\
  & ($\pi\Delta$)$_D$  &      28(1)  &      16(8)  &   \\
  & $\pi$N$^*$(1440)  &       4(1)  &       6(9)  &   \\
\\
P$_{31}(1750)$  & $\pi$N  &       6(6)  &       8(3)  &   \\
  & $\pi$N$^*$(1440)  &      83(1)  &      28(9)  &   \\
  & Flux  &      11(11)  &      64(9)  &   \\
\\
P$_{31}(1910)$  & $\pi$N  &      29(29)  &      23(8)  & 15-30  \\
  & $\pi$N$^*$(1440)  &      56(7)  &      67(10)  &   \\
  & Flux  &      15(15)  &      10(1)  &   \\
\\

P$_{33}(1232)$  & $\pi$N  &     100(1)  &     100(0)  & 98-100  \\
  & ($\pi\Delta$)$_P$  &       0(1)  &       0(0)  &   \\
  & $\pi$N$^*$(1440)  &       0(1)  &       0(0)  &   \\
\\
P$_{33}(1600)$  & $\pi$N  &      28(5)  &      12(2)  & 10-25  \\
  & ($\pi\Delta$)$_P$  &      59(10)  &      67(5)  & 40-70  \\
  & $\pi$N$^*$(1440)  &      13(4?1)  &      20(4)  & 10-35  \\

\\
P$_{33}(1920)$  & $\pi$N  &       5(61?)  &       2(2)  & 5-20  \\
  & ($\pi\Delta$)$_P$  &      41(2.9)  &      83(26)  &   \\
  & $\pi$N$^*$(1440)  &      53(8.2)  &      15(24)  &   \\
\\

D$_{33}(1700)$  & $\pi$N  &       5(1.6)  &      14(6)  & 10-20  \\
  & ($\rho_3$N)$_S$  &       1(1)  &       8(4)  & 5-20  \\
  & ($\pi\Delta$)$_D$  &       4(1)  &       4(3)  & 1-7  \\
  & ($\pi\Delta$)$_S$  &      90(1.7)  &      74(7)  & 25-50  \\
\\
D$_{35}(1930)$  & $\pi$N  &       9(8)  &      18(2)  & 5-20  \\
  & K$\Lambda$  &      91(11)  &   &   \\
  & Flux  &   &      82(2)  &   \\
\\
D$_{35}(2350)$  & $\pi$N  &       7(14)  &       2(0)  &   \\
  & K$\Lambda$  &      93(15)  &   &   \\
  & Flux  &   &      98(0)  &   \\
\\

F$_{35}(1752)$  & $\pi$N  &       0(1)  &       2(1)  &   \\
  & ($\rho_3$N)$_P$  &      60(60)  &      22(14)  &   \\
  & ($\pi\Delta$)$_F$  &      40(1)  &      48(16)  &   \\
  & ($\pi\Delta$)$_P$  &       0(1)  &      28(18)  &   \\
\\
F$_{35}(1905)$  & $\pi$N  &       9(2.2)  &      12(3)  & 5-15  \\
  & ($\rho_3$N)$_P$  &      24(1)  &      86(3)  & 0-60  \\
  & ($\pi\Delta$)$_F$  &      44(1)  &       0(1)  & 0-25  \\
  & ($\pi\Delta$)$_P$  &      23(1)  &       1(3)  &   \\
\\
F$_{37}(1950)$  & $\pi$N  &      44(1)  &      38(1)  & 35-40  \\
  & ($\rho_3$N)$_F$  &   &      43(1)  & 0-10  \\
  & ($\pi\Delta$)$_F$  &      36(1)  &      18(3)  & 20-30  \\
  & Flux  &      20(20)  &   &   \\
\\
\end{tabular}

\end{center}
\end{table}

\setcounter{table}{8}

\begin{table}
\label{tb:poles}
\caption{Pole positions.  The complex energy of the pole for each state 
is given along with the physical mass.}
\begin{tabular}{@{\extracolsep{\fill}}cccc}
Resonance  & Res Mass   &  Pole Position \\
           &  (MeV))   & (MeV)         \\
S$_{11}(1535)$ & 1545 & 1525 - 51 i\\
S$_{11}(1650)$ & 1693 & 1663 - 120 i\\
S$_{11}(2090)$ & 1822 & 1795 - 110 i\\
P$_{11}(1440)$ & 1479 & 1383 - 158 i\\
P$_{11}(1710)$ & 1699 & 1679 - 66 i\\
P$_{11}(2100)$ & 2083 & 1810 - 311 i\\
P$_{13}(1720)$ & 1716 & 1692 - 47 i\\
D$_{13}(1520)$ & 1520 & 1504 - 56 i\\
D$_{13}(1700)$ & 1729 & 1704 - 78 i\\
D$_{13}(2080)$ & 2002 & 1824 - 307 i\\
D$_{15}(1675)$ & 1687 & 1674 - 60 i\\
F$_{15}(1680)$ & 1679 & 1667 - 61 i\\
F$_{17}(1990)$ & 2311 & 2301 - 101 i\\
G$_{17}(2190)$ & 2168 & 2107 - 190 i\\

S$_{31}(1620)$ & 1633 & 1607 - 74 i\\
S$_{31}(1900)$ & 1798 & 1795 - 29 i\\
P$_{31}(1750)$ & 1721 & 1714 - 34 i\\
P$_{31}(1910)$ & 1995 & 1880 - 248 i\\
P$_{33}(1232)$ & 1234 & 1217 - 48 i\\
P$_{33}(1600)$ & 1687 & 1599 - 156 i\\
P$_{33}(1920)$ & 1889 & 1880 - 60 i\\
D$_{33}(1700)$ & 1732 & 1726 - 59 i\\
D$_{35}(1930)$ & 1932 & 1883 - 125 i\\
D$_{35}(2350)$ & 2459 & 2427 - 229 i\\
F$_{35}(1752)$ & 1724 & 1697 - 56 i\\
F$_{35}(1905)$ & 1873 & 1793 - 151 i\\
F$_{37}(1950)$ & 1936 & 1910 - 115 i\\
\end{tabular}

\end{table}


\begin{figure}
\caption{Schematic diagram for the Dyson equation iteration.} 
\label{fig:Dyson}
\end{figure}

\begin{figure}
\caption{Schematic diagram for the self energy ($\Sigma$) iteration.
Each resonance is allowed to couple to all open channels.}
\label{fig:Sigdiagram}
\end{figure}

\begin{figure}
\caption{$\phi(s)$ (channel propagator) distributions for the $\pi N$ and 
$\eta N$ channels in S$_{11}$ partial wave.  These functions are used
in the two channel-two resonance test case and in the final fits.} 
\label{fig:phi}
\end{figure}

\begin{figure}
\caption{One resonance-one channel test case $T$ matrix (real and 
imaginary parts) using parameters appropriate for the 
P$_{33}$(1232) resonance.  Solid dots are placed at the energy
of the resonance mass.}
\label{fig:p33}
\end{figure}

\begin{figure}
\caption{Self energy function, $\Sigma_{1535,1535}$ for the test case
where only $\pi N$ and $\eta N$ channels are included.}
\label{fig:S11sigma}
\end{figure}

\begin{figure}
\caption{$T$ matrices for two channel-two resonance test case.
The real and imaginary parts are shown for 
the $\pi N \rightarrow \pi N$ and  $\pi N \rightarrow \eta N$ reactions.
Solid dots are placed at the values of the resonance masses obtained
from these $T$ matrices.}
\label{fig:s11model}
\end{figure}

\begin{figure}
\caption{Best fit calculation for S$_{11}$ partial wave at the 
$\eta N$ threshold.  $T$ matrices for 
$\pi N \rightarrow \pi N$ and  $\pi N \rightarrow \rho_1 N$ are
shown.} 
\label{fig:S11cusp}
\end{figure}

\begin{figure}
\caption{$T$-matrices for S$_{11}$ partial wave for elastic scattering 
and final states of $\rho_1 N$, $\sigma N$, and $\pi N^*(1440)$ calculated 
with the full model using final fit parameters.  
The dotted line is calculated with only resonance couplings turned on and
the dashed line is calculated with only non-resonant couplings enabled.
The full calculation, which cannot be a sum of the dotted and dashed
lines because the resonant and non-resonant diagrams interfere,
is shown as a solid line.} 
\label{fig:S11rbf}
\end{figure}

\begin{figure}
\caption{$T$ matrices according to the best 
fit for the D$_{15}$ partial wave.  
The VPI partial wave amplitudes are shown by data points 
with error bars.  The KSU fit of Manley and Saleski is shown 
as dashed lines and
the fit of this work is shown as solid lines.}
\label{fig:D15}
\end{figure}

\begin{figure}
\caption{$T$ matrices for D$_{13}$ partial wave.  
The same labeling is used as in Fig.~\ref{fig:D15}.}
\label{fig:D13}
\end{figure}

\begin{figure}
\caption{$T$ matrices for S$_{11}$ partial wave.  
The same labeling is used as in Fig.~\ref{fig:D15}.}
\label{fig:S11}
\end{figure}

\begin{figure}
\caption{Total cross section for the 
$\pi^- p \rightarrow \eta n$ reaction.
The data was taken from sources in Ref.~[27], leaving out
the low energy Brown, {\it et al.} data. 
The solid line shows the total cross
section calculated with the final fit
result of this work using all partial waves.  
The contribution from the S$_{11}$ partial 
wave would be very similar to the full result.}
\label{fig:pietaxs}
\end{figure}

\begin{figure}
\caption{Pitt-ANL model calculation using final fit compared
         with the total cross section for the process
         $\pi^- p \rightarrow \pi^0 n$.  The fit was 
	 to $T$-matrix amplitudes derived in part from the data shown
         here.  Data shown is from the VPI SAID 
         database.} 
\label{fig:cxs_totxs}
\end{figure}

\begin{figure}
\caption{Pitt-ANL model calculation using final fit parameters compared
         to the differential cross section
         for the process $\pi^+ p \rightarrow \pi^+ p$.  Data from the
         VPI SAID database is shown.}
\label{fig:pipp_dsg}
\end{figure}

\begin{figure}
\caption{Pitt-ANL model calculation using final fit parameters
         of the polarization, P,
         for the process $\pi^- p \rightarrow \pi^- p$.  Data from the
         VPI SAID database is shown.}
\label{fig:pimp_pol}
\end{figure}

\title{References}

\end{document}